\documentclass[a4paper,11pt]{article}
\pdfoutput=1 
\usepackage{jheppub} 
\usepackage[T1]{fontenc}
\usepackage[italian,english]{babel}
\usepackage{hyperref}
\hypersetup{
	colorlinks=true,
	linkcolor=[rgb]{0.0,0.42,0.89},
	filecolor=magenta,      
	urlcolor=[rgb]{0.0,0.42,0.89},
	citecolor=[rgb]{0.64,0.0,0.0},
}
\usepackage{ifpdf}
\usepackage{subfigure}
\usepackage{gensymb}
\usepackage{amssymb}
\usepackage{amsfonts}
\usepackage{slashed}
\usepackage{epsf}
\usepackage{rotating}
\usepackage{graphicx}
\usepackage{amsmath}
\usepackage{fancyhdr}
\usepackage{lineno}
\usepackage{babel, blindtext}
\usepackage{graphics}
\usepackage{color}
\usepackage{physics}
\usepackage{multirow}
\usepackage{pstricks}
\usepackage{xcolor}
\usepackage{multirow}
\usepackage{hhline}
\usepackage{cancel}
\usepackage{framed}
\usepackage{mathtools}
\usepackage{diagbox}
\usepackage{float}
\usepackage{orcidlink}
\usepackage{ulem} 
\usepackage[lmargin=1.7in,bmargin=0.2in,footskip=0.5in,total={6.9in,9.5in}]{geometry}

\newcommand{\nn}{\nonumber}
\newcommand{\lsim}{\mathrel{\mathop{\kern 0pt \rlap
			{\raise.2ex\hbox{$<$}}}
		\lower.9ex\hbox{\kern-.190em $\sim$}}}
\newcommand{\gsim}{\mathrel{\mathop{\kern 0pt \rlap
			{\raise.2ex\hbox{$>$}}}
		\lower.9ex\hbox{\kern-.190em $\sim$}}}

\newcommand{\be}{\begin{equation}}
	\newcommand{\ee}{\end{equation}}
\newcommand{\bea}{\begin{eqnarray}}
	\newcommand{\eea}{\end{eqnarray}}

\title{\boldmath Finite-temperature stability from doublet inflation field with right-handed neutrinos}

\author[a]{Seong Chan Park,\orcidlink{0000-0003-0176-4355}}
\author[b,c]{Shilpa Jangid\, \orcidlink{0000-0001-6307-1234}}

\affiliation[a]{	Department of Physics,                                                                                           IPAP, Lab for Dark Universe, Yonsei University, 50 Yonsei-ro, Seodaemun-gu,
	Seoul 03722, Korea}
\affiliation[b]{
	Asia Pacific Center for Theoretical Physics (APCTP) - Headquarters San 31,
	Hyoja-dong, Nam-gu, Pohang 790-784, Korea}
\affiliation[c]{Shiv Nadar IoE Deemed to be University, Gautam Buddha Nagar, Uttar Pradesh, 201314, India}

\emailAdd{sc.park@yonsei.ac.kr, shilpajangid123@gmail.com}

\preprint{}

\abstract{ We study the augmentation of the Standard Model (SM) with another $SU(2)$ Higgs doublet and right-handed neutrinos. The second Higgs doublet ($\Phi_2$) is defined to be odd under the $Z_2$ symmetry, and hence, the lightest stable neutral particle from the additional doublet becomes the cold dark matter candidate. The right-handed neutrino field coupled to the Higgs field provides non-zero mass for the neutrinos. The inert doublet field coupled non-minimally to gravity as $\zeta_2 \Phi_2^\dagger \Phi_2 R$ also acts as an inflaton field. The inflationary bounds restrict the interaction couplings as $\lambda_2/\zeta_2^2 \approx 4\times 10^{-10}$. After inflation ends, the scalar bosonic degrees of freedom from the inert doublet can contribute to the electroweak phase transition. The strongly first-order phase transition bound, i.e., $\frac{\phi_{+}(T_c)}{T_c} \geq 1.0$ restricts the bare mass parameter of the additional doublet to $m_{22}=400.0$ GeV, demanding GUT scale perturbative unitarity  for $Y_N=0.01$. The right-handed neutrino effect is studied here only through the running of Yukawa coupling, i.e. $Y_N$. The running of Yukawa coupling for different values at the electroweak scale restricts the perturbative unitarity at different scales, and effects the strength of phase transition.  The increase in $Y_N$ reduces the strength of phase transition, and it is no longer satisfied even for vanishing bare mass parameter. The Planck scale perturbative unitarity allows for the first-order phase transition, $\frac{\phi_{+}(T_c)}{T_c} \geq 0.6$, until $m_{22}=70.0$ GeV for $Y_N=0.01$, and none of the mass values satisfies the first-order phase transition for $Y_N=0.4$. The thermal corrections also affect the probability of tunneling from the false vacuum to the true vacuum, and hence, the finite temperature stability of the electroweak vacuum has been studied including the finite-temperature effects. }


\begin{document}

\maketitle
\flushbottom

\section{Introduction} 
The discovery of the Higgs boson at the Large Hadron Collider (LHC) as predicted by the Standard Model (SM) was a moment of triumph for particle physicists. The Standard Model (SM) came across as a complete theory with this discovery. Still, there are various theoretical and experimental evidences for several scenarios that are not accommodated by the SM. The absence of inflation, the cold dark matter (DM) candidate, non-zero neutrino mass, first-order phase transition (FOPT), and the stability of the electroweak vacuum until Planck scale are a few of them. The non-zero neutrino mass was confirmed by the neutrino oscillation experiment, and the two-zero texture stands out an attractive framework to predict the unknown parameters of the PMNS matrix \cite{Biswas:2025woq, Jangid:2025ded}. Then, there is firm evidence from the cosmic microwave background (CMB) for both the inflation, and the dark matter (DM). The theory of inflation being so successful explains different cosmological problems such as the well known horizon problem and homogeneity \cite{Guth:1980zm}. There are a lot of defined theories explaining inflation. Higgs inflation is the minimal amongst them, with just one parameter, $\zeta_1$, explaining the non-minimal coupling with gravity. This model suffers from Planck scale vacuum stability as the Higgs field self coupling, $\lambda_1$, turns negative around $10^9-10^{10}$ GeV \cite{Sher:1988mj,Buttazzo:2013uya,Degrassi:2012ry} and secondly, it suffers from non-unitarity too. Out of the experimental bounds on the inflationary parameters such as the spectral index, the tensor to scalar ratio, and the spectral power spectrum from Planck \cite{Planck:2015sxf} and WMAP7 \cite{WMAP:2010qai}, the parameter $\zeta_1$ is restricted to $\sim 10^{4}$ \cite{Lerner:2009na}, which breaks the unitarity at scales around $m_{Pl}/\zeta_1 \sim 10^{13}$ GeV. These shortcomings require the extension of the SM. The minimal extension of the SM can be with a gauge singlet, which can act as an inflaton field, and the Higgs field acts as a portal to reheat the Universe \cite{Lerner:2011ge}. The inflationary potential can be either a chaotic one or a Starobinsky one; the chaotic inflation models have power law potentials, and the Starobinsky ones have exponential potentials \cite{Linde:1983gd,Linde:2014nna}. The minimal extension of the SM with a gauge singlet gives Starobinsky's kind of potential with the best fit to inflationary parameters like the spectral index. The second minimal extension can be with another $SU(2)$ Higgs doublet, which will also give a Starobinsky kind of potential.  \\  

Similar to the inflation theory, the excess of matter over antimatter also requires new scalar degrees of freedom beyond the Standard Model. There are several ways to generate this baryon asymmetry of the Universe (BAU), and the baryon asymmetry generation from the electroweak phase transition is an interesting aspect. The out of equilibrium dynamics, which is the necessary criteria for explaining the excess of matter over antimatter generated by electroweak baryogenesis (EWBG), occurs during the strongly first-order phase transition  \cite{Trodden:1998ym,Cohen:1993nk,Morrissey:2012db,Baker:2021zsf,Cline:2020jre,Jangid:2023jya} via bubble nucleation \cite{PhysRevD.30.2212,Gavela:1994dt,Huet:1994jb}. But this possibility remains unfulfilled in the Standard Model (SM). The electroweak phase transition is not first-order \cite{Aoki:1999fi} in the case of SM for the Higgs boson mass larger than $80.0$ GeV \cite{Kajantie:1996mn,Kajantie:1996qd,Csikor:1998eu}, which is not at all consistent with the experimentally observed Higgs boson mass of 125.5 GeV \cite{Kajantie:1995kf}. The strongly first-order phase transition is also interesting because of the possibility of generating gravitational wave (GW) signatures \cite{Hindmarsh:2015qta}.  The recent observation of gravitational waves (GW) from black-hole mergers validated such observations \cite{LIGOScientific:2016aoc,LIGOScientific:2016dsl}, and there is a possibility for accessibility of the frequency range generated by the first-order phase transition at the electroweak (EW) scale by space-based detectors like eLISA \cite{eLISA:2013xep}. 
The possibility for the singlet extension of the SM explaining both inflation and the electroweak phase transition has been studied in detail (separately \cite{Lerner:2009xg,Cline:2012hg, Alanne:2014bra} and combined \cite{Choubey:2017hsq}). The next minimal extension is with another Higgs doublet with right-handed neutrinos, which can explain both inflation and the electroweak phase transition and has not been explored yet. The possibility of the first-order phase transition \cite{Benincasa:2022elt,Fabian:2020hny,Majumdar:2020vdd,Blinov:2015vma,Borah:2012pu,Fabian:2020hny} and inflation in the inert doublet \cite{Choubey:2017hsq} model have already been studied separately.\\

There is also a good estimate for the distribution of the DM in and around our galaxy and the Universe, which SM cannot accommodate. The scalar extensions with a singlet or a doublet employed with a discrete $Z_2$ symmetry can provide a stable neutral lightest particle which can act as the DM candidate. The extension with a singlet field or doublet field can act as an inflaton field, and later can freeze-out as the DM candidate. The motivation for extension of the inert doublet with right-handed neutrino will also provide non-zero neutrino mass \cite{Jangid:2020dqh} alongside inflation, first-order electroweak phase transition, and a cold DM candidate, which is another shortcoming of the SM. The Yukawa couplings are also crucial in effecting the phase transition dynamics by restricting the allowed values from Planck scale perturbativity. Lastly, the reheating of the Universe occurs at the end of inflation via the interactions of the inert doublet with the vector gauge bosons and the Higgs.  \\

As discussed above, the self-quartic coupling of the SM, $\lambda_1$, turns negative around $10^{9}-10^{10}$ GeV, and within the uncertainity of the top mass, we lie in the metastable vacuua \cite{Buttazzo:2013uya,Degrassi:2012ry} which is another theoretical motivation for additional scalar degrees of freedom.   
These additional scalar degrees of freedom from the second Higgs doublet contributes positively to the effective quartic coupling and later, to the stability of the EW vacuum. The instability of the electroweak vacuum has already been studied in detail \cite{Espinosa:2007qp,Espinosa:2015qea,Kobakhidze:2013tn,Enqvist:2013kaa,Fairbairn:2014zia,Enqvist:2014bua,Kobakhidze:2014xda,Herranen:2014cua,Kamada:2014ufa,Shkerin:2015exa} from the cosmological perspective. The quantum fluctuations because of the thermal corrections may force the Higgs field to tunnel down to the true vacuum during inflation. This scenario can even take place before the inflation ends, but it is not possible to realize this when the reheating temperature is sufficiently large after the end of inflation. Therefore, the Universe is identified by a sufficiently high temperature after inflation ends \cite{Espinosa:2015qea}. In consequence, the stability of the electroweak vacuum is affected by the thermal corrections \cite{Isidori:2001bm, DelleRose:2015bpo,PhysRevD.44.3620,Jangid:2023lny}, as the tunneling from the false vacuum to the true vacuum of the potential is triggered \cite{Coleman:1977py,Callan:1977pt,Anderson:1990aa,Espinosa:1995se}, and hence, cannot be neglected. Here, as we are considering the thermal corrections to the effective potential, it will be intriguing to examine the state for the stability of the EW vacuum both at the zero temperature and the finite temperature.\\

The detailed layout of the article is as follows; the detailed description of the model with field definitions and the mass expressions are given in \autoref{model}. The inflationary dynamics and the bounds from the spectral index and the scalar to tensor ratio are computed in \autoref{inf}. The allowed parameter space is tested for the strongly first-order phase transition in \autoref{EWPT}. The details for the stability of the electroweak vacuum at finite-temperature for different values of Yukawa couplings are given in \autoref{FT}. The relevant two-loop $\beta$-functions for the gauge couplings and the quartic couplings are given in \autoref{beta}, and the expressions for the two-loop corrections to the thermal masses using Dimensional reduction approach are given in \autoref{dim}.

\section{Model setup}\label{model}
We consider the minimal standard model (SM) extended with additional $SU(2)$ Higgs doublet and the right-handed neutrinos named "Type-I seesaw". The additional doublet is assigned to be  odd under the $Z_2$ symmetry, while the other fields, including the right-handed neutrino and the SM fields, are even under the discrete $Z_2$ symmetry. The second doublet coupled non-minimally to gravity acts as the inflaton field, and the interactions with the Higgs field will effect the behavior of the scalar potential at different temperatures. We discuss the effect of additional degrees of freedom on the phase transition dynamics together with the inflationary bounds. The field definition of the additional multiplets is as follows;
\bea
Z_2:  \Phi_1 \rightarrow \Phi_1, \Phi_2 \rightarrow -\Phi_2, N_i \rightarrow N_i,
\eea
where,
\begin{center}
	$	\Phi_1
	= \left(\begin{array}{c}
		G^+   \\
		\frac{1}{\sqrt{2}}(v_h+\rho_1+i G^0)  \end{array}\right) $, \qquad \qquad
	$\Phi_2 = \left(
	\begin{array}{c}
		 H^{+} \\
	\frac{H^0+i A^0}{\sqrt{2}}  \\
	\end{array}
	\right)$,
\end{center}
where $\Phi_2$ being odd under the $Z_2$ symmetry does not participate in the electroweak symmetry breaking (EWSB), and freezes-out as the cold dark matter (DM) candidate after the inflation ends. The most relevant CP conserving, renormalizable potential for the inert doublet model \cite{LopezHonorez:2010eeh,Gustafsson:2010zz,Treesukrat:2019ahh,Goudelis:2013uca,LopezHonorez:2012zz,Tytgat:2007cv,LopezHonorez:2007wm,LopezHonorez:2006gr} is written as given below;
\begin{eqnarray}\label{eq:2.2}
	\rm  V_{scalar} &&= m_{11}^2\Phi_1^\dagger \Phi_1 + m_{22}^2\Phi_2^\dagger\Phi_2 + \lambda_1(\Phi_1^\dagger \Phi_1)^2 + \lambda_2(\Phi_2^\dagger \Phi_2)^2 +
	\lambda_3(\Phi_1^\dagger \Phi_1)(\Phi_2^\dagger \Phi_2) \nn \\ &&+ \lambda_4(\Phi_1^\dagger \Phi_2)(\Phi_2^\dagger \Phi_1) + [\frac{\lambda_5}{2}((\Phi_1^\dagger \Phi_2)^2) + h.c].  \label{Eq:2.3}
\end{eqnarray}
The scalar masses after electroweak symmetry breaking (EWSB) are given as;
\begin{eqnarray}\label{mass1}
	 M_{h}^2 &= &2\lambda_1 v_h^2 \nn\\
	 M_{H^0}^2 &=& \frac{1}{2}(2m_{22}^2+v_h^2(\lambda_3+ \lambda_4+\lambda_5))\nn\\
	 M_{A^0}^2 &=& \frac{1}{2}(2m_{22}^2+v_h^2(\lambda_3+\lambda_4-\lambda_5))\nn\\
	 M_{H^\pm}^2 &=& m_{22}^2+\frac{1}{2}v_h^2
	\lambda_3.
\end{eqnarray} 
Depending upon the sign of $\lambda_5$, either $M_{A^0}$ or $M_{H^0}$ can be the dark matter candidate (DM). Here, we consider $\lambda_5$ to be negative and $M_{H^0}$ to be the cold DM candidate. The Yukawa interactions additional to the Standard Model (SM) are given as follows;
 \begin{eqnarray}
	\mathcal{L}_{\rm I} \ = \ i \overline{N}_{i} \slashed{\partial} N_{i}- \left(Y_{N_{ij} } \overline{L}_i \widetilde{\Phi}_1 N_{j} + \frac{1}{2}\overline{N}_{i}^c M_{N_i} N_{i} +  h.c. \right)\, ,
\end{eqnarray}
 where $Y_{N}$ is the 3$\times$3
 Yukawa matrix, and $M_N$ is the 3$\times$3 diagonal mass matrix for RHNs. 
 
 The additional doublet which is odd under the $Z_2$ symmetry, acts as the inflaton field, enters the thermal equilibrium with the rest of the plasma and evolves as radiation. Later, the inert doublet becomes non-relativistic with the decrease in temperature, and its evolution is governed by the Boltzmann equation. The detailed computation for the inflationary dynamics with the inert doublet is discussed in \autoref{inf}.
 
 \section{Inflation}\label{inf}
 The Standard Model (SM) extended with the second Higgs doublet is considered to be coupled non-minimally to gravity. The additional Higgs doublet is assigned to be odd under the $Z_2$ symmetry and hence, there are no couplings with the fermions. The action for this specific model is written as;
 \bea\label{eq:3.1}
 S_J= \int d^4 x \sqrt{-g}\Big[-\frac{1}{2}M_{Pl}^2R - |D_{\mu}\Phi_1|^2 -|D_{\mu}\Phi_2|^2 -V_{scalar}(\Phi_1, \Phi_2) -2\zeta_1 \Phi_1^\dagger \Phi_1 R -2\zeta_2 \Phi_2^\dagger \Phi_2 R\Big],
 \eea
 where $D_\mu$ is the covariant derivative denoting the couplings to the gauge bosons, and it will reduce to the normal derivative $(D_{\mu} \rightarrow\partial_{\mu})$ during inflation in the absence of other fields. $M_{Pl} =1/\sqrt{8\pi G}\approx 2.4\times 10^{18}~{\rm GeV}$  is  the reduced Planck mass, $R$ is the Ricci scalar, and $\zeta_1$ and $\zeta_2$ are the corresponding dimensionless couplings of the doublets to the gravity. These particular terms come from the quantum effects at Planck scales. The doublet fields, in general, can be rewritten in the component form as given below:
 \begin{center}
 	$	\Phi_1
 	= \frac{1}{\sqrt{2}}\left(\begin{array}{c}
 		\chi   \\
 		h \end{array}\right)$, \qquad \qquad
 	$\Phi_2 =\frac{1}{\sqrt{2}} \left(
 	\begin{array}{c}
 		q \\
 		\varphi e^{i\theta} \\
 	\end{array} \right)$.
 \end{center}
Here, vacuum expectation value for the Higgs field is neglected as the electroweak symmetry stays intact at the inflationary scales. In order to ensure the second doublet to be the inflation field, we request the following conditions: 
\begin{itemize}
\item $V(\Phi_1, \Phi_2) \approx \lambda_2 (\Phi_2^\dagger \Phi_2)^2 = \frac{\lambda_2}{4}\varphi^4 $  during inflation
\item $ \zeta_1 |\Phi_1|^2  \ll \zeta_2 |\Phi_2|^2  \approx \zeta_2 \varphi^2/2$
\end{itemize}
then inflation takes place along $\varphi$-direction (inert, neutral direction).

The action given in \autoref{eq:3.1} is in Jordon frame, and we need to get rid of the terms quadratically coupled to gravity in order to recover the canonically normalized gravity in Einstein frame. After Weyl transformation, the relevant part of the action is now given as follows;
\bea
S_E = \int d^4 x \sqrt{- \widetilde{g}}\Big[-\frac{1}{2}M^2_{Pl}\widetilde{R} -\frac{1}{2} G_{ij}\widetilde{g}^{\mu \nu }\partial_{\mu}\phi_1 \partial_\nu \phi_j -U(h, q, \varphi, \theta)\Big],
\eea
where, $\phi = \{ \chi, h, q, \varphi, \theta\}$ and  
\bea
\widetilde{g}_{\mu\nu} & = & \Omega^2 g_{\mu \nu},\\
\Omega^2 & = &  1+ \frac{\zeta_1}{M^2_{Pl}}(\chi^2 + h^2) + \frac{\zeta_2}{M^2_{Pl}}(q^2+\varphi^2) \approx 1+\zeta_2 \frac{\varphi^2}{M_{Pl}^2},\\
G_{ij} & = & \frac{1}{\Omega^2}\delta_{ij} + \frac{3}{2} \frac{M^2_{Pl}}{\Omega^4}\frac{\partial \Omega^2}{\partial \phi_i}\frac{\partial \Omega^2}{\partial \phi_j}, \\
U & = & \frac{V}{\Omega^4} \approx \frac{\lambda_2 \varphi^4}{4 (1+\zeta_2 \varphi^2/M_{Pl}^2)^2}.
\eea
Keeping this in mind, the only relevant field that contributes during inflation comes from the ${\mathbb{Z}_2}$-odd doublet, and after some further simplification, the potential  along $\varphi$-direction is as follows:
\bea\label{eq:3.7}
U_e \approx \frac{\lambda_2 M^4_{Pl}}{4 \zeta_2^2}\Big[ 1- \exp\Big(-\sqrt{\frac{2}{3}} \frac{X}{M_{pl}}\Big)\Big],~X=\sqrt{\frac{3}{2}}M_{Pl}\log(\Omega^2) .
\eea
This potential guarantees the successful inflation~\cite{Bezrukov:2007ep, Park:2008hz} : 
\begin{align}
n_s \approx 0.965, ~r \approx 0.003
\end{align}
as long as $\lambda_2/\zeta_2^2 \approx 4.4 \times 10^{-10}$ for the Planck normalization~\cite{Planck:2013jfk}.
Reheating by $\varphi$ decays to the standard model sector is accomplished via kinetic mixings and the Higgs quartic interactions. 
The estimated reheating temperature is 
\bea
T_r \simeq \Big(\frac{30 \rho_r}{\pi^2 g_*}^{1/4}\Big) \simeq 10 ^{14} \rm GeV,
\eea 
where $g_{*}=116.0$ is the relativistic degrees of freedom in plasma, including the SM particles, three $Z_2$ even SM-singlet fermions, and four additional scalar degrees of freedom from the second Higgs doublet~\cite{Choubey:2017hsq}.

\section{Electroweak phase transition}\label{EWPT}
The phase transition from the symmetric phase at higher temperature to the broken phase at lower temperatures requires a finite-temperature analysis. The tree-level potential given in \autoref{eq:2.2} will get an additional contribution from the zero-temperature Coleman-Weinberg one-loop effective potential  and  finite-temperature one-loop potential. The second Higgs doublet is inert, and hence, the mass expressions will be rewritten in terms of background field for the SM-Higgs field. The full finite-temperature one-loop effective potential is written as given below;
\bea
V_1(\Phi_1, \Phi_2, T)= V_0^{\rm eff}(\Phi_1,\Phi_2) + V_{\rm 1-loop}^{\rm CW}(\widetilde{m}_i^2) + V_{\rm 1-loop}^{T \neq 0} (\widetilde{m}_i^2).
\eea
where, 
$V_0^{\rm eff}=V_{scalar}$ is the tree-level potential at zero temperature in terms of background fields, $V_{\rm 1-loop}^{\rm CW}(\widetilde{m}_i^2)$ is the one-loop Coleman-Weinberg effective potential at zero-temperature, and $V_{\rm 1-loop}^{T \neq 0} (\widetilde{m}_i^2)$ is the one-loop effective potential at a finite-temperature. The corresponding expressions are given as follows \cite{Coleman};
\bea\label{eq:5.2}
V_{1-loop}^{\rm CW} & = & \frac{1}{(64\pi)^2}\sum_{i=B,F}(-1)^{F_i} n_i \hat{m_i}^4\Big[\log\Big(\frac{\widetilde{m}_i^2}{\mu^2}\Big)-k_i\Big], \\
V_{1-loop}^{T \neq 0} &  = & \frac{T^4}{(2\pi)^2}\sum_{i=B,F} (-1)^{F_i} n_i J_{B/F}\Big(\frac{\widetilde{m}_i^2}{T^2}\Big).
\eea
where, $\widetilde{m}_i^2$ are the thermally corrected masses, which include contributions from the Daisy corrections resuming hard thermal loops;
\bea
\widetilde{m}_i^2 = \widetilde{m}_i^2(h_1; T) = \hat{m}_i^2(h_1) + \Pi_i T^2,
\eea
where $\Pi_i's$ are the Daisy coefficients, which are non-zero only for the bosonic fields. The quantum corrections from the Coleman-Weinberg one-loop effective potential are scale dependent, and they will effect the phase transition dynamics if the values at the scale $\mu$ are very different from the values chosen at the electroweak scale. $h_1$ is the background field for the SM Higgs doublet, and the mass expressions $\hat{m}_i^2(h_1)$ in terms of the background field are given as;
\begin{center}
	$\mathcal{M}^2_{CP_{even}} 
	=  \left(\begin{array}{cc}
	m_{11}^2+3\lambda_1 h_1^2 +\frac{(\lambda_3+\lambda_4+\lambda_5)h_2^2}{2}& (\lambda_3+\lambda_4+\lambda_5)h_1 h_2  \\
		(\lambda_3+\lambda_4+\lambda_5)h_1 h_2  & m_{22}^2 + \frac{(\lambda_3+\lambda_4+\lambda_5)h_1^2}{2} + 3\lambda_2 h_2^2
		 \end{array}\right) $,
\end{center}
\begin{center}
	$\mathcal{M}^2_{CP_{odd}} 
	=  \left(\begin{array}{cc}
		m_{11}^2+\lambda_1 h_1^2 +\frac{(\lambda_3+\lambda_4-\lambda_5)h_2^2}{2}& \lambda_5 h_1 h_2  \\
		\lambda_5 h_1 h_2  & m_{22}^2 + \frac{(\lambda_3+\lambda_4-\lambda_5)h_1^2}{2} + \lambda_2 h_2^2
	\end{array}\right) $,
\end{center}
\begin{center}
	$\mathcal{M}^2_{\pm} 
	=  \left(\begin{array}{cc}
		m_{11}^2+\lambda_1 h_1^2 +\frac{\lambda_3}{2}h_2^2& \frac{(\lambda_4+\lambda_5)}{2} h_1 h_2  \\
		\frac{(\lambda_4+\lambda_5)}{2} h_1 h_2 & m_{22}^2 + \frac{
		\lambda_3h_1^2}{2} + \lambda_2 h_2^2
	\end{array}\right) $,
\end{center}
\bea
\hat{m}_{G^0}^2 = \lambda_h h_1^2 - m_h^2, \qquad \hat{m}_{W}^2 = \frac{g_2^2}{4} h_1^2, \qquad \hat{m}_{Z}^2 = \frac{g_2^2 + g_1^2}{4}h_1^2, \qquad \hat{m}_t^2 =\frac{y_t^2}{2}h_1^2, \nn
\eea
where, $\hat{m}_{G_0}, \hat{m}_{W}^2, \hat{m}_Z^2, \hat{m}_t^2$ are the field dependent mass expressions for the Goldstone bosons, the SM gauge bosons, and the top quark, respectively \cite{Jangid:2023lny}.
The finite-temperature corrections would be the same for all degrees of freedom from the same multiplet. The transverse degrees of freedom for the gauge bosons are protected from the gauge symmetry, and the corresponding thermal corrections are zero, i.e., $\Pi_{W_T}=\Pi_{Z_T}=\Pi_{\gamma_T}=0$. 
The expressions for the finite-temperature corrections to these masses are as follows;
 \bea
\Pi_h & = & \Big(\frac{g_1^2+3g_2^2}{16} + \frac{\lambda_h}{2}+ \frac{y_t^2}{4}+ \frac{2\lambda_3+\lambda_4}{12}\Big)T^2, \nn \\
\Pi_{G^0,G^{\pm}} & = & \Big(\frac{g_1^2+3g_2^2}{16} + \frac{\lambda_h}{2}+ \frac{y_t^2}{4}+ \frac{2\lambda_3+\lambda_4}{12} \Big)T^2, \nn \\
\Pi_{H^0,A^0,H^{\pm}} & = & \Big(  \frac{\lambda_{2}}{2} + \frac{2\lambda_3+\lambda_4}{12}\Big)T^2, \nn \\
\Pi_{W_L} & = & 2g_2^2 T^2, \nn \\
\Pi_{W_T} & = & \Pi_{Z_T}=\Pi_{\gamma_T} =0, \nn \\
\widetilde{m}_{Z_L}^2 & = & \frac{1}{2}\hat{m}_Z^2 +\frac{g_2^2}{\cos^2\theta_W}T^2 + \delta, \nn \\
\widetilde{m}_{\gamma_L}^2 & = & \frac{1}{2}\hat{m}_Z^2 +\frac{g_2^2}{\cos^2\theta_W}T^2 - \delta, \nn \\
\eea
where, $\delta$ is given as;
\bea
\delta^2 = \Big(\frac{1}{2}\hat{m}_Z^2 + \frac{g_2^2}{\cos^2\theta_W}T^2\Big)^2 -g_1^2 g_2^2 T^2(h_1^2 + 4 T^2).
\eea

The two-loop $\beta$-functions are used for computing the vacuum stability and the perturbative unitarity bounds and are computed using SARAH \cite{Staub:2013tta}. The two-loop expressions for the running of the quartic couplings and the gauge couplings are provided in \autoref{beta}. Since, we are using two-loop $\beta$-functions and one-loop effective finite temperature potential, the effective potential at finite temperature has residual scale dependence at $\mathcal{O}(g^4)$. The cancellation of this scale dependence
at $\mathcal{O}(g^4)$ requires the inclusion of two-loop thermal masses to bare masses for Higgs, and the doublet, i.e. $m_{11}$ and $m_{22}$, respectively. The dimensional reduction technique is a
systematic approach required at high temperature to the resummations done order-by-order in power of couplings. In short, in order to cancel the scale dependence, we need to include the two-loop corrections to the thermal masses. This is achieved by reducing the theory to 3dEFT, where all the parameters in the potential become scale dependent, using dimensional reduction technique \cite{Gorda:2018hvi,Andersen:1998br}. The detailed expressions for the dimensionally reduced parameters are given in \autoref{dim}. The further analysis for the electroweak phase transition is performed taking into account the dimensionally reduced parameters. The stability of the potential at tree-level is ensured with the following constraints on the quartic couplings \cite{Branco:2011iw};
\bea
\lambda_1> 0,\quad \lambda_2>0, \quad 2\sqrt{\lambda_1\lambda_2}+\lambda_3>0, \quad 2\sqrt{\lambda_1\lambda_2}+\lambda_3+\lambda_4-2|\lambda_5|>0,
\eea
and the perturbativity constraints for the dimensional couplings are given as follows \cite{Jangid:2020dqh,Bandyopadhyay:2020djh};
\bea
|\lambda_i| \leq 4\pi, \quad  |g_j|\leq 4\pi, \quad  |Y_k| \leq \sqrt{4\pi},
\eea
where $i \in \{1,2,3,4,5\}, j\in \{1,2,3\}, k \in \{u,d,e,N\}$ for the quartic, gauge, and the Yukawa couplings. The electroweak phase transition using one-loop $\beta$-functions has been studied in the context of inert doublet \cite{Blinov:2015vma}. This analysis is extended to two-loop $\beta$-functions and the SM input parameters at the EW scale are given in \autoref{tab:SMint}. 

\begin{table}[h!]
	\centering
	\begin{tabular}{|c|c|c|c|c|}
		\hline
		$g_1$&$g_2$& $g_3$&$Y_u^{33}$&$\lambda_h$\\
		\hline
		0.46256&0.64779&1.1666&0.93690&0.12604\\
		\hline
	\end{tabular}
	\caption{EW scale values of the parameters for the RG evolution using two-loop $\beta$- functions.}
	\label{tab:SMint}
\end{table}

\begin{figure}[H]
	\begin{center}
		\mbox{\subfigure[$\mu=246.0$ GeV ]{\includegraphics[width=0.48\linewidth,angle=-0]{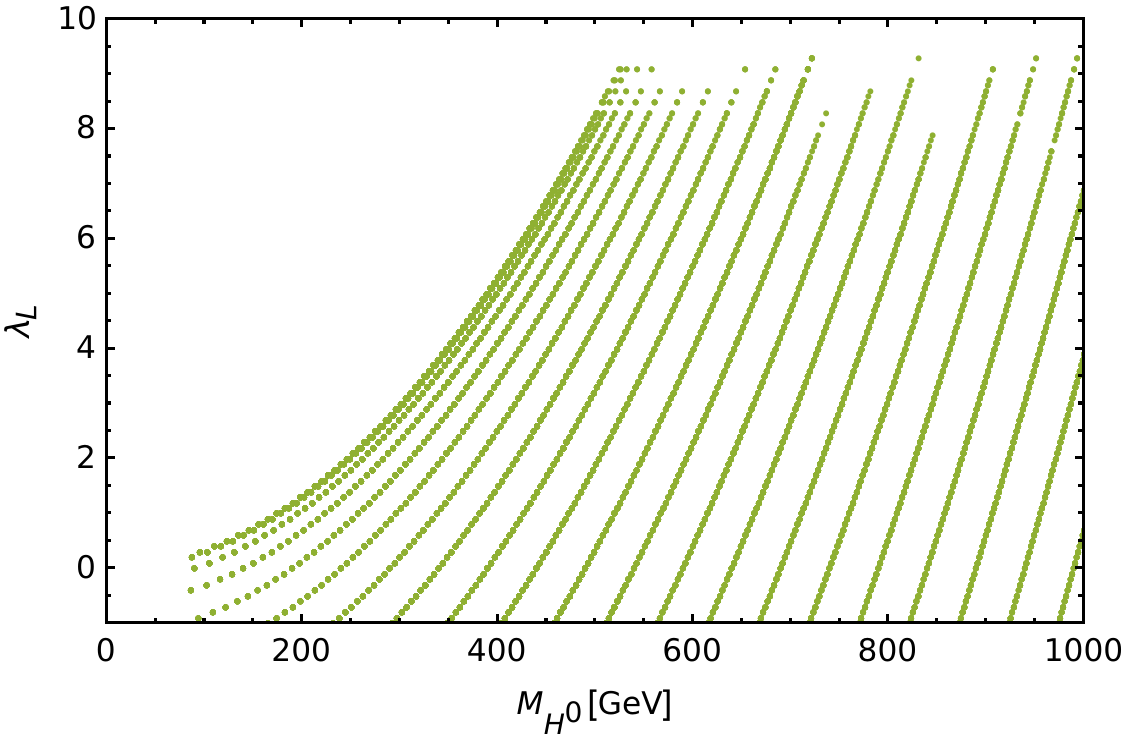}\label{f1a}}
			\subfigure[$\mu=246.0$ GeV ]{\includegraphics[width=0.5\linewidth,angle=-0]{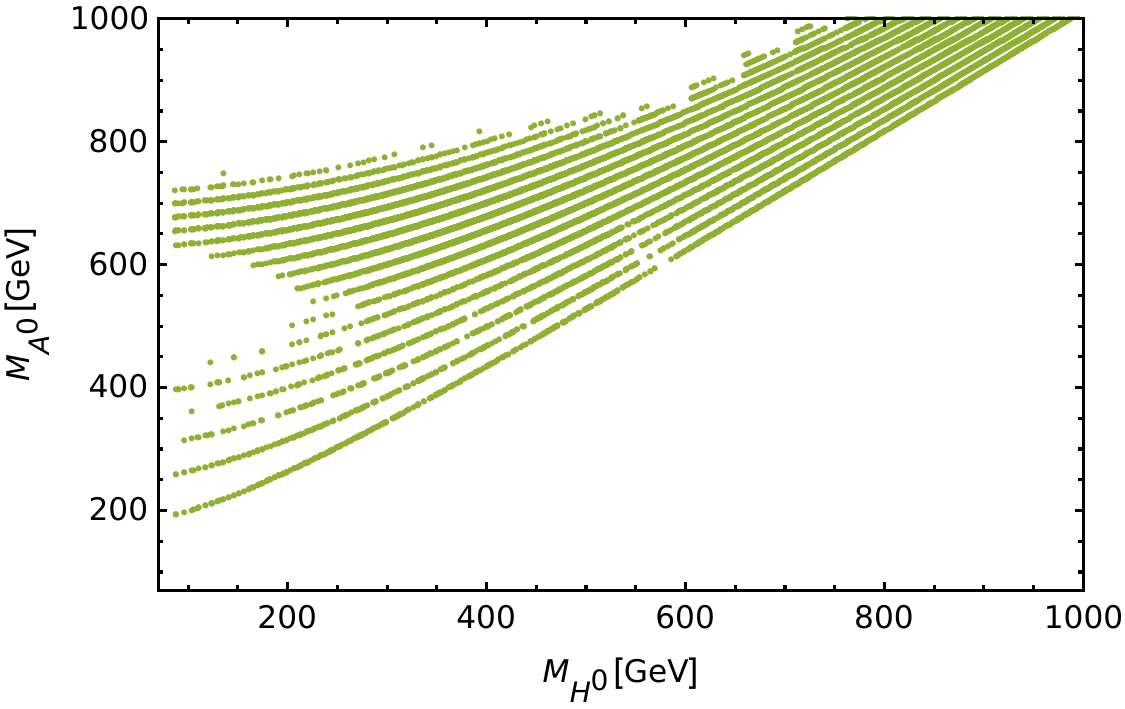}\label{f2a}}.}\caption{(a) denotes the DM mass $M_{H^0}$  variation in GeV with the interaction quartic coupling $\lambda_L=(\lambda_3+\lambda_4+\lambda_5)$ and the mass splitting between the DM mass $M_{H^0}$ and $M_{A^0}$ is given in (b). The green points describes the allowed parameter space from strongly first-order phase transition. The running of the couplings is considered from the EW scale $m_t$, to $\mu=246$ GeV and the couplings are constrained to be perturbative till the scale $\mu=246$ GeV. The SM input parameters at two-loop level chosen at the EW scale are given in \autoref{tab:SMint}. }\label{fig:246}
	\end{center}
\end{figure}

The scale dependence $\mu$ in \autoref{eq:5.2} becomes crucial when these input parameters at the EW scale vary with the scale. In that case, it becomes important to take into account the running of the couplings to minimize the scale dependence. The scale $\mu$ is chosen to be equal to 246 GeV, and the couplings are chosen to be perturbative at this particular scale. The variation of the DM mass in GeV with the interaction quartic coupling $\lambda_L=(\lambda_3+\lambda_4+\lambda_5)$ is plotted in \autoref{fig:246}(a), and the mass splitting in GeV between the DM mass $(M_{H^0})$ and $M_{A^0}$ is shown by \autoref{fig:246}(b). For $\mu=246$ GeV, the input values are fed at the EW ($m_t$) scale, and the running is considered till 246 GeV, demanding perturbative unitarity till 246 GeV. The green points denote the criteria for strongly first-order phase transition, i.e., $\phi_{+}(T_c)/T_c \geq 1$. The strongly first-order phase transition is satisfied for all possible mass values for such high values of the interaction coupling. The value of $Y_N$ is chosen to be 0.01, but the allowed parameter space would remain unaltered even for higher values of $Y_N$.

\begin{figure}[H]
	\begin{center}
			\mbox{\subfigure[$\mu=10^{15}$ GeV $, Y_N=0.01$ ]{\includegraphics[width=0.48\linewidth,angle=-0]{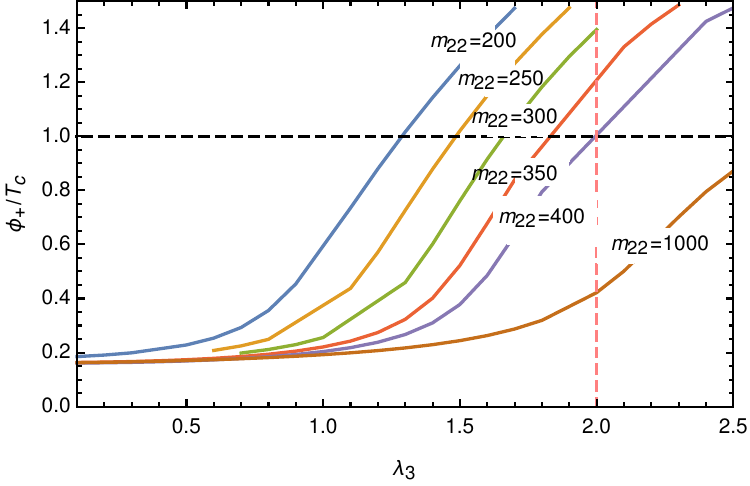}\label{f1a}}
			\subfigure[$\mu=10^{15}$ GeV  $,Y_N=0.4$ ]{\includegraphics[width=0.48\linewidth,angle=-0]{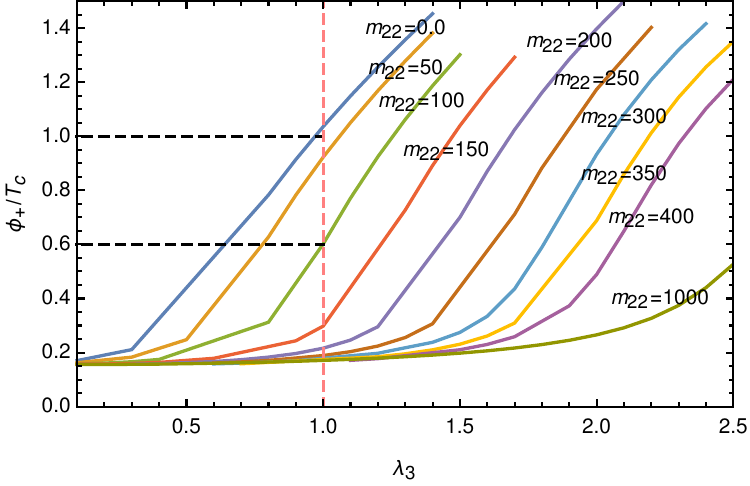}\label{f2a}}.}
	\caption{(a) denotes the variation of the interaction coupling $\lambda_3$ with the strength of phase transition for different values of the doublet mass parameter $(m_{22})$. The perturbative scale is chosen till $\mu=10^{15}$ GeV and the dashed pink line correspond to the maximum possible allowed value of the interaction quartic coupling $\lambda_3$ at the EW scale for $Y_N=0.01$ and $Y_N=0.4$ in \autoref{fig:06}(a) and (b), respectively. The upper dashed black line correspond to the criteria for strongly first-order phase transition, i.e, $\phi_{+}(T_c)/T_c \sim 1$ and the lower one denotes the first-order phase transition, i.e, $\phi_{+}(T_c)/T_c \sim 0.6$. }\label{fig:06}
	\end{center}
\end{figure}
The positive effect in the case of an inert doublet is quite large because of the larger number of degrees of freedom, and hence, the quartic couplings would be restricted to very low values from the perturbative unitarity bound as we go higher and higher in scale.  Therefore, the variation of the interaction quartic coupling $\lambda_3$ with the strength of electroweak phase transition is shown in \autoref{fig:06} for different values of the doublet mass parameter. In this case, perturbative unitarity is demanded up to $10^{15}$ GeV. The upper horizontal black dashed line delineates the criteria of a strongly first-order phase transition, i.e, $\phi_{+}(T_c)/T_c \geq 1$, and the lower one denotes the first-order phase transition, i.e., $\phi_{+}(T_c)/T_c \geq 0.6$. The pink dashed line defines the maximum possible allowed value of the interaction quartic coupling $\lambda_3$ allowed from the perturbative unitarity till $10^{15}$ GeV. The maximum possible allowed value for the interaction quartic coupling from GUT scale perturbativity is $\lambda_3=2.0$ for $\lambda_4=1.2$ and $\lambda_5=-1.4$ for $Y_N=0.01$, as demonstrated in \autoref{fig:06}(a). It is clear from \autoref{fig:06}(a) that the strongly first-order phase transition is satisfied until $m_{22}=400.0$ GeV. The situation changes drastically as the value of $Y_N$ is increased to 0.4 in \autoref{fig:06}(b) because of the positive contribution from the $\rm Tr(Y_N^{\dagger}Y_N)$ term at two-loop in the running of the quartic couplings other than the SM self quartic coupling. This positive contribution reduces the maximum allowed value for the interaction quartic coupling $\lambda_3$ to 1.0 from GUT scale perturbativity for $\lambda_4=0.14$ and $\lambda_5=-0.18$, and now the strongly first-order phase transition is achieved only for vanishing doublet mass parameter. The first-order phase transition is still achieved until $m_{22}=100.0$ GeV. This puts bound on the allowed value for the $Y_N$ from the strongly first-order phase or first-order phase transition.

\begin{figure}[H]
	\begin{center}
		\mbox{\subfigure[$\mu=10^{19}$ GeV $, Y_N=0.01$ ]{\includegraphics[width=0.48\linewidth,angle=-0]{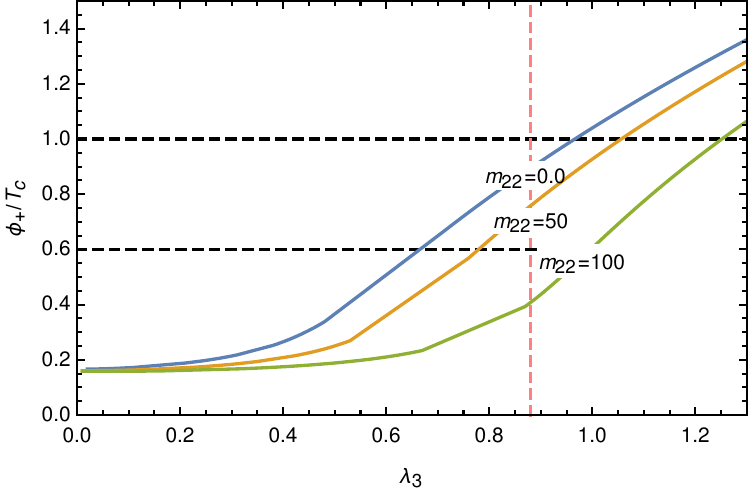}\label{f1a}}
			\subfigure[$\mu=10^{19}$ GeV  $,Y_N=0.4$ ]{\includegraphics[width=0.48\linewidth,angle=-0]{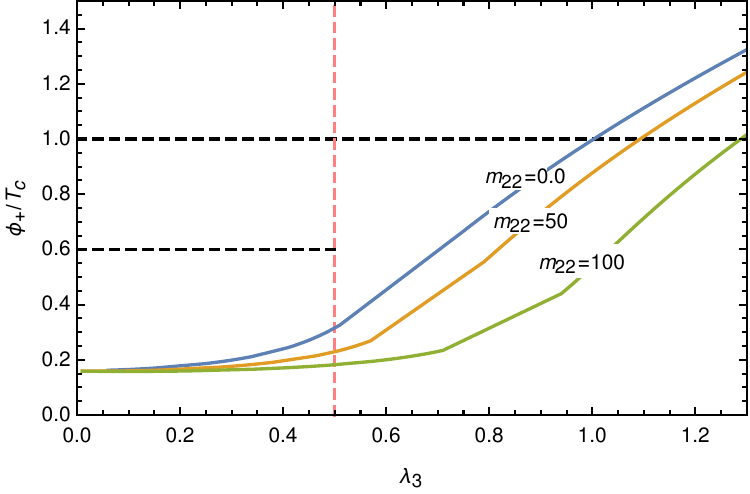}\label{f2a}}.}\caption{(a) Variation of the interaction quartic coupling $\lambda_3$ with the strength of phase transition $\phi_{+}(T_c)/T_c$ for $Y_N=0.01$ and $Y_N=0.4$ in (b). The perturbative scale is chosen to be $\mu=10^{19}$ GeV and the mass parameter is constrained to $m_{22}\lsim70$ GeV from the first-order phase transition as shown by (a). }\label{fig:19}
	\end{center}
\end{figure}

After discussing the bounds from the GUT scale perturbativity, the perturbative scale is extended to $\mu=10^{19}$ GeV in \autoref{fig:19}. For perturbative scale up to Planck scale $10^{19}$ GeV, the parameter space is too much constrained. The variation of the interaction quartic coupling $\lambda_3$ with the strength of phase transition is given in \autoref{fig:19}(a)-(b) for two different values of $Y_N$. This would put a bound on the mass parameter $m_{22}\lsim 70$ GeV from the first-order phase transition as shown in \autoref{fig:19}(a) for $Y_N=0.01$. The pink dashed line defines the maximum possible allowed value of the interaction quartic coupling $(\lambda_3)$ at the EW scale from Planck scale perturbativity for $\lambda_4=0.14$ and $\lambda_5=-0.18$. As the value of $Y_N$ is increased to 0.4, the quartic couplings are restricted to further lower values because of the positive contribution from the term $\rm Tr(Y_N^{\dagger}Y_N)$ at two-loop, i.e., $\lambda_3=0.50$ as denoted by the dashed pink line in \autoref{fig:19}(b) for $\lambda_4=0.0$ and $\lambda_5=-0.10$. For such high value of $Y_N$, none of the values of the mass parameter $m_{22}$ satisfy the criteria for first-order phase transition. This favors the lower values of $Y_N$ for phase transition to be of first-order. The important thing is that the values allowed in \autoref{fig:19}(a) are also allowed from the inflation bounds.

The discussion of the bounds for the electroweak phase transition ends here. It is worthwhile to investigate whether or not this electroweak vacuum is stable because of the phase transition from the symmetric phase at higher temperatures to the broken phase at lower temperatures. As a result, we concentrate on the electroweak stability of the vacuum at both zero and finite temperatures in the next section. 

\section{Finite temperature stability}\label{FT}
The Planck scale stability of the electroweak vacuum is the theoretical drawback and it is not achieved in the case of Standard Model within the uncertainity of top quark mass.  The tunneling likelihood from the electroweak vacuum to the second minimum, which is present at higher field values, is further triggered by the thermal corrections. Therefore, testing the electroweak vacuum stability from zero to a high enough temperature is an interesting task. 
By examining the bounce solution of the classical equation of motion and accounting for the classical potential, $V=\frac{\lambda_1}{4}h_1^4$, the decay lifespan of the electroweak vacuum is computed \cite{Kobakhidze:2013tn,DelleRose:2015bpo}. According to Coleman \cite{Coleman:1977py,Callan:1977pt}, the equation of motion at zero temperature is expressed as follows;
\bea \label{eq:6.1}
\frac{d^2h_1}{dr^2} + \frac{3}{r}\frac{dh_1}{d r}=\frac{dV(h_1)}{dh_1}, \qquad  \lim_{r \rightarrow \infty}h_1(r)=0, \qquad \frac{dh_1}{dr}|_{r=0} =0,
\eea
with $r= |\vec{r}|$. The $O(4)$ spherically symmetric solution in \autoref{eq:6.1} gives euclidean action as follows:
\bea
S_E[h_1(r)]= 2\pi^2 \int_{0}^{\infty} dr r^3 \Big[\frac{1}{2}\Big(\frac{dh_1}{dr}\Big)^2 + V(h_1)\Big].
\eea
The following formula is used to calculate the bounce solutions for the classical equation of motion in \autoref{eq:6.1}:
\bea\label{eq:6.3}
h_{1B}(r)=\frac{8}{|\lambda_1|} \frac{R}{R^2+r^2},  \qquad \qquad \qquad S_E[h_{1b}(r)]=\frac{8\pi^2}{3 |\lambda_1|},
\eea
where, $R$ is the arbitary scale defining the size of the bounce $(0 < R < \infty)$. The  appearance of this arbitary parameter in \autoref{eq:6.3} is because of our approximations that the potential considered above is scale invariant. The scale invariance of the potential leads to an infinite set of bounce solutions giving the same value for action. But this scale invariance of the classical tree-level potential is broken by the quantum corrections. This implies that the bounce solutions which used to give same action at the semi-classical level for different chosen values of $R$ now gives one particular one-loop action as $S[h_{1b}(r)] \sim \frac{8 \pi^2}{(3 |\lambda(1/R)|)}$. Considering the effective potential as $V_{eff}=\frac{\lambda_{eff}}{4}h_1^4$, the bounce solution in \autoref{eq:6.3} now becomes,
\bea\label{eq:5.4}
h_{1B}(r)=\frac{8}{|\lambda_{eff}(\mu)|} \frac{R}{R^2+r^2},  \qquad \qquad \qquad S_E[h_{1b}(r)]=\frac{8\pi^2}{3 |\lambda_{eff}(\mu)|}.
\eea
with $\lambda_{eff}$ is the effective quartic coupling which incorporates contributions from all the particles coupled to the Higgs field. Now, for this particular action, there is just one specific value of $R$, defined as $R_M$ will saturate the path integral. As the variation of effective quartic coupling $\lambda_{eff}$ is considered with the energy scale, $\lambda_{eff}(\mu)$ approaches zero, and eventually approaches negative values, indicating the occurrence of an instability. Here, the action is minimized by $\beta_{\lambda_{eff}}(\mu)=0$, where $R_M\sim 1/\mu$ represents the bounce size and $\mu$ is the renormalization scale. It is simple to use \autoref{eq:5.4} to draw the bounce profile after determining the scale $\mu$.  

\subsection{Case-I, $(m_t = 172.69+ \delta m + 0.3), \ \lambda_i=0.01, \ Y_N=0.01$ }
\begin{figure}[h!]
	\begin{center}
		\mbox{\subfigure[$\lambda_i = 0.01$, $ Y_N=0.01$ ]{\includegraphics[width=0.48\linewidth,angle=-0]{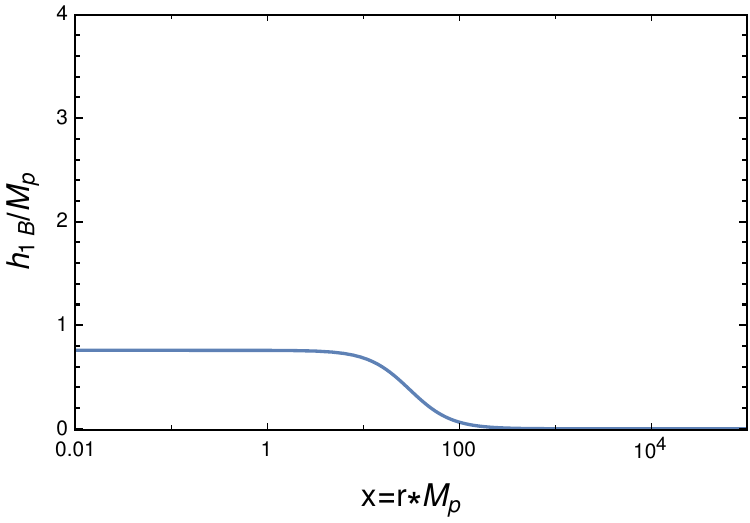}\label{f1a}}}
		\caption{The variation of the field, $h_1$ with the euclidean distance, $r$ for quartic couplings $\lambda_i=0.01$ with $i \in \{2,3,4,5\}$ and $Y_N=0.01$. The field $h_1$, and the euclidean action are both rescaled with the Planck mass, $M_P=1.22 \times 10^{19}$ GeV. The value of the bounce solution at the center comes out to be $h_{1B}(0)=9.638 \times 10^{18}$ GeV. }\label{fig:bounce2}
	\end{center}
\end{figure}

The stability of the electroweak vacuum is sensitive to the top quark mass, but this theoretical uncertainity is hard to quantify. The direct measurements extract the top-quark mass from the kinematics of $t\bar{t}$ events with average value $172.69 \pm 0.3$ from LHC and Tevatron runs \cite{Workman:2022ynf}. The combined uncertainities in the top-quark mass $\delta m$ is approximated as $\pm 1$ GeV. This subsection is devoted to the analysis with $m_t = 172.69+ \delta m + 0.3$. The detailed analysis for the profile of the bounce with top quark mass, i.e., $m_t =172.69+ \delta m + 0.3 $, the quartic couplings as $\lambda_i=0.01$ where $i \in {2,3,4,5}$ and the Yukawa coupling as $Y_N=0.01$ is given in \autoref{fig:bounce2}. Both the four-dimensional euclidean distance, $r$, and the field $h_{1B}$ are scaled using the Planck mass $M_P=1.22 \times 10^{19}$ GeV. The value of $R_M$ comes out to be $30.64 \ M_P^{-1}$. Using this particular value of $R_M$ in the effective quartic coupling in \autoref{eq:5.4}, the value of the bounce at the center, i.e., $h_{1B}(0)= 9.638 \times 10^{18}$ GeV at zero temperature.

The thermally corrected mass expressions and the full one-loop potential including the thermal-loops using resummation is given in \autoref{EWPT}. Still for the completeness the expressions for the one-loop potential are given as follows;
\bea
V_{\rm 1-loop}^{T\neq 0} = \sum_{i=W,G^0,Z,h,\Delta,\Phi_2} \frac{n_i T^4}{ 2\pi^2}J_B\Big(\frac{\hat{m}_i^2}{T^2}\Big) + \frac{n_tT^4}{2 \pi^2} J_F\Big(\frac{\hat{m}_t^2}{T^2}\Big), \label{eq:6.5}\\
V_{\rm ring}^{T\neq 0} = \sum_{i=W_L,G^0,Z_L,\gamma_L,h,\Delta,\Phi_2} \frac{n_iT^4}{12 \pi}\Big\{\Big[\frac{\hat{m}_i^2}{T^2}\Big]^{3/2}-\Big[\frac{\widetilde{m}_i^2 }{T^2}\Big]^{3/2}\Big\}\label{eq:6.6},
\eea
where the degrees of freedom are $n_i's$, the Debye masses are $\widetilde{m}_i^2$, and the field dependent masses are $\hat{m}_i^2$. It is noteworthy that only the longitudinal degrees of freedom exhibit non-zero thermal corrections for the gauge bosons; the fermionic degrees of freedom exhibit zero thermal corrections. The spline functions $J_{B,F}$ have the following specified expressions:
\bea
J_{B,F}(x^2) = \int_{0}^{\infty} dy y^2 \log(1 \mp e^{-\sqrt{y^2+x^2}}).
\eea

Now, the bounce solution in \autoref{eq:5.4} is modified by the thermal corrections as described below \cite{Anderson:1990aa,PhysRevD.44.3620,Espinosa:1995se};
\bea \label{eq:6.7}
\frac{d^2h_1}{dr^2} + \frac{2}{r}\frac{dh_1}{d r}=\frac{dV_{eff}(h_1,T)}{dh_1}, \qquad  \lim_{r \rightarrow \infty}h_1(r)=0, \qquad \frac{dh_1}{dr}|_{r=0} =0,
\eea
with $r= |\vec{r}|$ and $V_{eff}(h_1,T)$ is full one-loop thermally corrected potential in terms of background field, $V_{eff}(h_1,T)= V_{\rm scalar} + V_{\rm 1-loop}^{\rm CW}+ V_{\rm 1-loop}^{T \neq 0} + V_{\rm ring}^{T \neq 0}$. The euclidean action for this $O(3)$ spherically symmetric solution is as follows;
\bea
S_3[h_1(r)]= 4\pi \int_{0}^{\infty} dr r^2 \Big[\frac{1}{2}\Big(\frac{dh_1}{dr}\Big)^2 + V_{eff}(h_1,T)\Big].
\eea

\begin{figure}[h!]
	\begin{center}
		\mbox{\subfigure[$\lambda_i = 0.01$, $ Y_N=0.01$ ]{\includegraphics[width=0.48\linewidth,angle=-0]{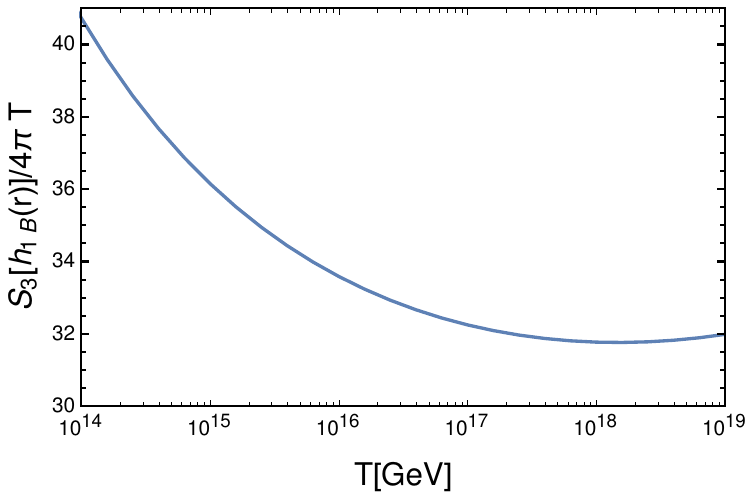}\label{f1a}}}
		\caption{The variation of the action with temperature $T$ in GeV.}\label{fig:action2}
	\end{center}
\end{figure}

In the high-temperature limit, the full thermally corrected one-loop effective potential can be rewritten as follows:
\bea\label{eq:6.10}
V_{eff}(h_1,T) \simeq \frac{\lambda_{eff}}{4}h_1^4 + \frac{1}{2}\eta^2 h_1^2 T^2 + \rm const
\eea
where the constant terms are those which are unaffected of the field $h_1$, and hence, are neglected. The coefficient for the quadratic terms in field and the temperature dependence is given as;
\bea
\eta^2 & = & \frac{1}{12}\Big(\frac{3}{4}g_1^2 + \frac{9}{4} g_2^2 + 3 y_t^2 + 6 \lambda_{1} + 2 \lambda_3 +\lambda_4 \Big)-\frac{\sqrt{2}}{32\pi}(g_1^3+3g_2^3) \nn \\
&& - \frac{\sqrt{3}}{16 \pi}\lambda_{1} \sqrt{3 g_1^2+9 g_  2^2 +24 \lambda_{1}+12 y_t^2 + 8 \lambda_3 +4 \lambda_4}  -  \frac{1}{16 \sqrt{3}\pi} (2\lambda_3 +\lambda_4)\sqrt{6\lambda_2 + (2\lambda_3 +\lambda_4)},      
\eea
where \autoref{eq:6.5}'s finite temperature one-loop potential provides the first term, and \autoref{eq:6.6}'s resummation provides the second and third terms. The bounce solution at finite temperature can be directly determined as $S_3[h_{1B}(r)] \simeq -(6.015)\pi \eta / \lambda_{eff}T$ using the effective potential expression found in \autoref{eq:6.10}. It is important to keep in mind that all of the couplings in the effective quartic coupling and $\eta$ are scale dependent, and that the running values are considered for all of these parameters. \autoref{fig:action2} shows how this action varies with temperature, $T$ in GeV.

\begin{figure}[h!]
	\begin{center}
		\mbox{\subfigure[$\lambda_i = 0.01$, $ Y_N=0.01$ ]{\includegraphics[width=0.48\linewidth,angle=-0]{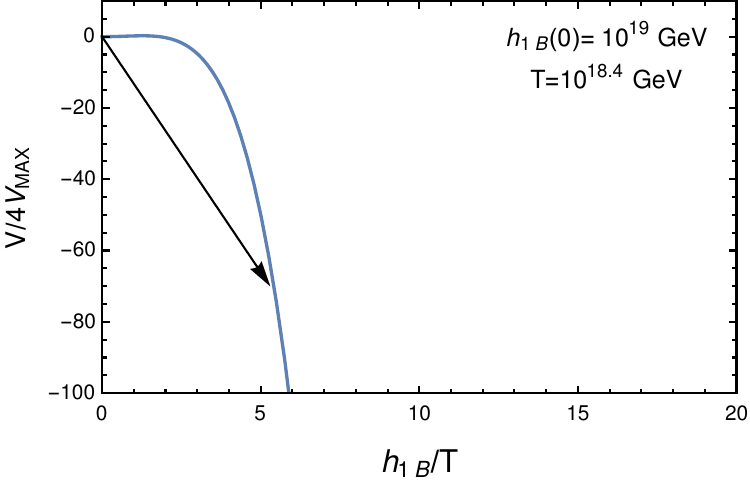}\label{f1a}}
			\subfigure[$\lambda_i = 0.01$, $ Y_N=0.01$ ]{\includegraphics[width=0.48\linewidth,angle=-0]{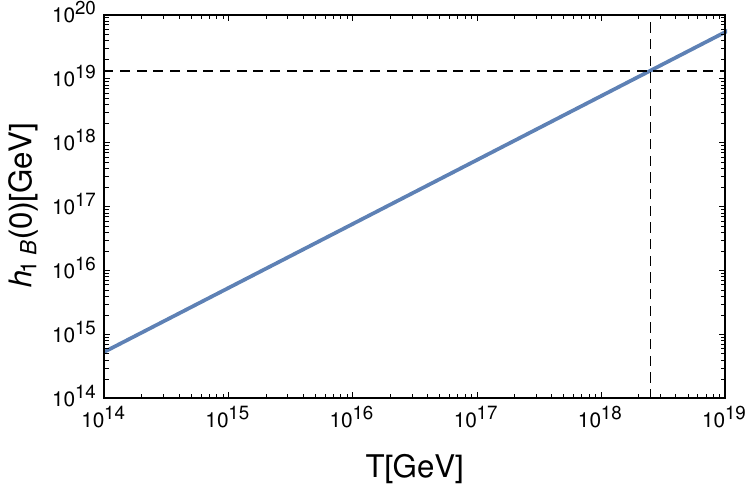}\label{f2a}}.}\caption{(a) The variation of the effective potential $V/4V_{MAX}$ with $h_{1B}(0)/T$ for one particular temperature $T=10^{18.4}$ which leads to the center of bounce till Planck scale. The ratio $h_{1B}(0)/T$ always comes out to be $ \sim 5.3$ throughout the analysis which is shown by the trip of the arrow. (b) shows the variation of the center of bounce with different temperatures and the dashed black line denotes the center of bounce at Planck scale and the corresponding value of the temperature. }\label{fig:phib0idm34}
	\end{center}
\end{figure}

After computing the bounce variation, the value of bounce at the center $h_{1B}(0) = 10^{19}$ GeV is achieved at finite temperature $T=10^{18.4}$ GeV which is shown by the tip of the arrow in \autoref{fig:phib0idm34}(a). The value of the bounce at the center at different temperatures is shown in \autoref{fig:phib0idm34}(b). The ratio $h_{1B}(0)/T$ always comes out to be $ \sim 5.3$ throughout the analysis. This will give a cut-off for the maximum temperature if we look for the validity of the theory till Planck scale. The value of bounce solution at the center, $h_{1B}(0)$ comes out to be order of Planck scale at temperature $T=10^{18.4}$ GeV. The dashed black line denotes the center of bounce at Planck scale and the corresponding value of the temperature. Hence, we can choose the $T_{\rm cut-off}$ to be $10^{18.4}$ GeV which we will need later on for computing the total transition probability from the false vacuum to the true vacuum.

It is simple to assess the differential decay probability of the nucleating bubble at a specific temperature $T$ after computing the action;
\bea \label{eq:5.12}
\frac{dP}{d \ln T} \simeq \Gamma(T) \frac{M_P}{T^2}\Big(\frac{\tau_U T_0}{T}\Big)^3,
\eea
with $\Gamma(T)\simeq T^4 \Big\{ \frac{S_3[h_{1B}(r)]}{2 \pi T}\Big\}^{3/2} e^{-S_3[h_{1B}(r)]/T}$ is the vacuum decay rate per unit volume at a fixed temperature, $T$. $\tau_U$ is the age of the Universe and $T_0 \simeq 2.35 \times 10^{-4} $ eV \cite{DelleRose:2015bpo}. The differential decay probability given in \autoref{eq:5.12} is true only in a radiation dominated Universe.

\begin{figure}[H]
	\begin{center}
		\mbox{\subfigure[$\lambda_i = 0.01$, $ Y_N=0.01$ ]{\includegraphics[width=0.48\linewidth,angle=-0]{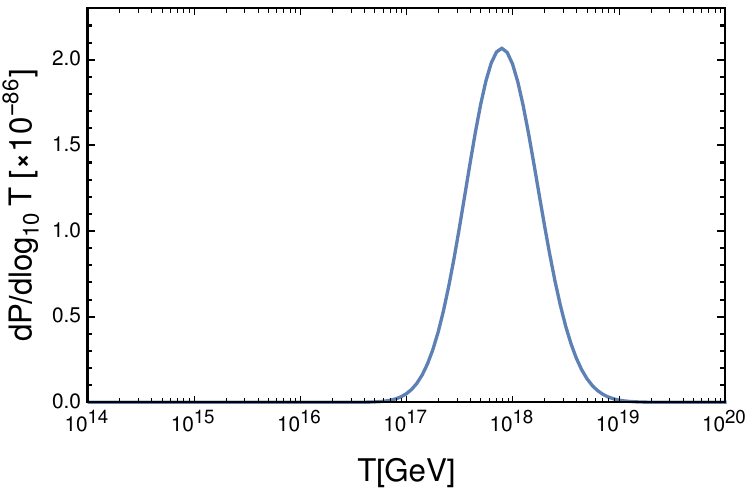}\label{f1a}}
			\subfigure[$\lambda_i = 0.01$, $ Y_N=0.01$ ]{\includegraphics[width=0.48\linewidth,angle=-0]{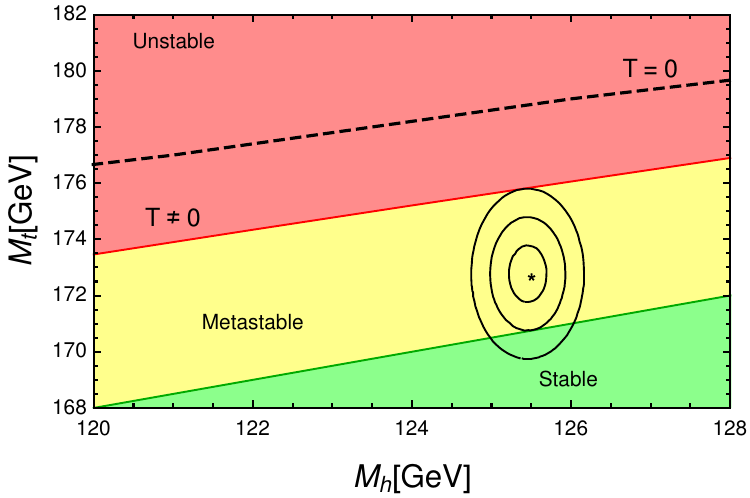}\label{f2a}}.}\caption{(a) Variation of the differential probability with the temperature in GeV for $m_t = 172.69 +  \delta m + 0.3 $ and $\lambda_i=0.01, \ Y_N=0.01$, (b) phase space plot for the Higgs mass vs top quark mass. The green and the yellow region denotes the stable and the metastable region, respectively. The red line and the black dashed line corresponds to the instability at finite-temperature and the zero temperature. }\label{fig:diffprob2}
	\end{center}
\end{figure}

By integrating the differential decay probability in \autoref{eq:5.12}, the total integrated probability may now be calculated directly as follows:
\bea\label{eq:5.13}
P(T_{\rm cut-off}) = \int_{0}^{T_{\rm cut-off}} \frac{dP(T')}{dT'}dT',
\eea
with $T_{\rm cut-off}$ is the maximum cut-off temperature which is computed by checking the validity till a particular cut-off scale as $\Lambda$ when the bounce $h_{1B}(0)$ touches the Planck scale. The profile for the differential decay probability with the temperature in GeV is given in \autoref{fig:diffprob2}(a). The peak of the differential decay probability is around $ \sim 10^{18.0}$ GeV and the total integrated probability is computed using \autoref{eq:5.13}. This decay probability from the false vacuua to the true vacuum can be used to separate the regions of stability, metastability and the instability. 

The criteria for $\lambda_{eff} > 0$ corresponds to the stable region while the instability region at zero temperature actually corresponds to tunneling probability $p= max_R \frac{\tau_U^4}{R^4} exp[-\frac{8\pi^2}{3|\lambda_{eff}(\mu)|}]>1$, with $\tau_U$ is defined as the age of the Universe (denoted by dashed black line the red region).
In case of metastability region (yellow), the effective quartic coupling, $\lambda_{eff}$ becomes negative below the Planck scale and the decay probability from the false vacuum to the true vacuum is actually $<1$. 
The instability at finite temperature denoted by red line corresponds to $P>1$ which is given by \autoref{eq:5.13}. The stable, metastable and the
unstable regions are plotted in \autoref{fig:diffprob2}(b) for $m_t = 172.69+ \delta m + 0.3 $ and $\lambda_i=0.01, \ Y_N=0.01$. The black dot corresponds to the central values for the Higgs mass and top quark mass and the contours corresponds to the $1 \sigma, \ 2 \sigma$ and $3 \sigma$ uncertainity. The thermal fluctuations increase the chance for tunneling from the EW minima to another second minima and the instability bound gets more stringent at finite tempearture in comparison to the zero temperature.

\subsection{Case-II, $(m_t = 172.69), \ \lambda_i=0.01, \ Y_N=0.01$ }
\begin{figure}[h!]
	\begin{center}
		\mbox{\subfigure[$\lambda_i = 0.01$, $ Y_N=0.01$ ]{\includegraphics[width=0.48\linewidth,angle=-0]{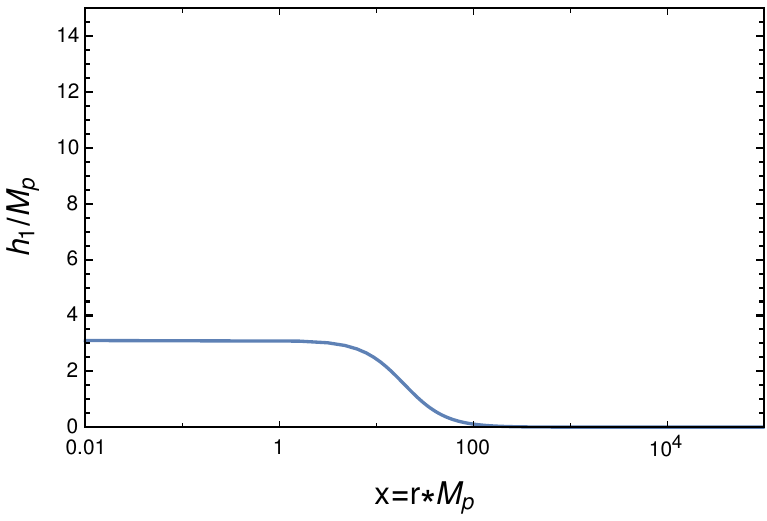}\label{f1a}}}
		\caption{The variation of the field, $h_1$ with the euclidean distance $r$ for quartic couplings $\lambda_i=0.01$ with $i \in \{2,3,4,5\}$. The field $h_1$ and the euclidean action are both rescaled with the Planck mass, $M_P=1.22 \times 10^{19}$ GeV. The value of the bounce solution at the center comes out to be $h_{1B}(0)=3.78 \times 10^{19}$ GeV. }\label{fig:bounce1}
	\end{center}
\end{figure}

Now, the top-quark mass is reduced and so the corresponding top-quark Yukawa coupling. Both the field profile $h_1$ and the four-dimensional euclidean distance $r$, which are scaled using the Planck mass $M_P=1.22 \times 10^{19}$ GeV, will be affected by this. The value of $R_M$ comes out to be 19.36 $ Mp^{-1}$ for quartic couplings $\lambda_{i}=0.01$, where $i \in \{2,3,4,5\}$. Using this value of $R_M$ and $\mu = 1/R_M$, the bounce can be plotted as given in \autoref{fig:bounce1}(a). The value of bounce at the center, i.e., $h_{1B}(0)$ at zero temperature is $3.78 \times 10^{19}$ GeV. The top quark coupling is lower as compared to the previous case and hence, the value of the effective quartic coupling becomes negative and saturates at lower scale. Therefore, the value of the bounce at the center is more according to \autoref{eq:5.4}.

\begin{figure}[h!]
	\begin{center}
		\mbox{\subfigure[$\lambda_i = 0.01$, $ Y_N=0.01$ ]{\includegraphics[width=0.48\linewidth,angle=-0]{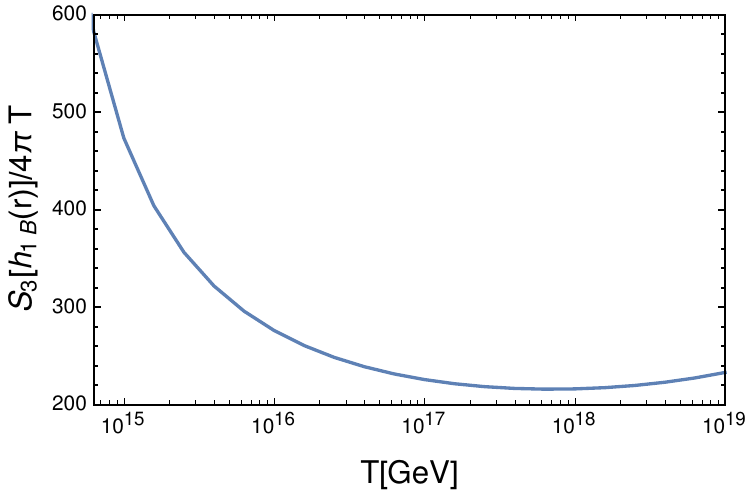}\label{f1a}}}
		\caption{The variation of the action with temperature $T$ in GeV.}\label{fig:action1}
	\end{center}
\end{figure}

\begin{figure}[H]
	\begin{center}
		\mbox{\subfigure[$\lambda_i = 0.01$, $ Y_N=0.01$ ]{\includegraphics[width=0.48\linewidth,angle=-0]{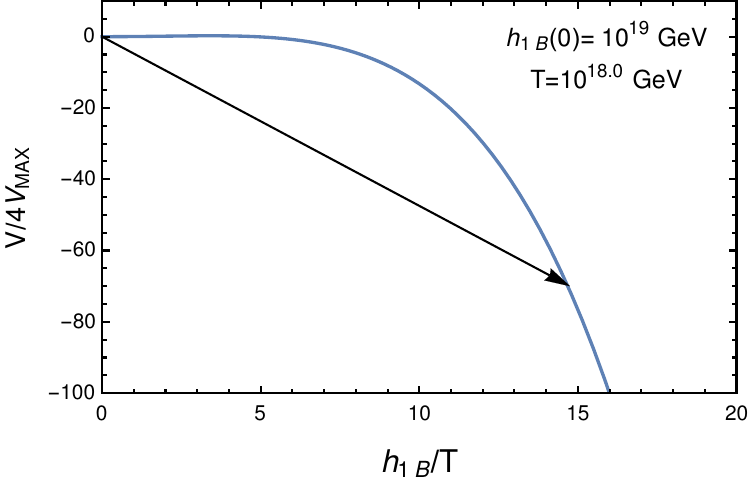}\label{f1a}}
			\subfigure[$\lambda_i = 0.01$, $ Y_N=0.01$ ]{\includegraphics[width=0.48\linewidth,angle=-0]{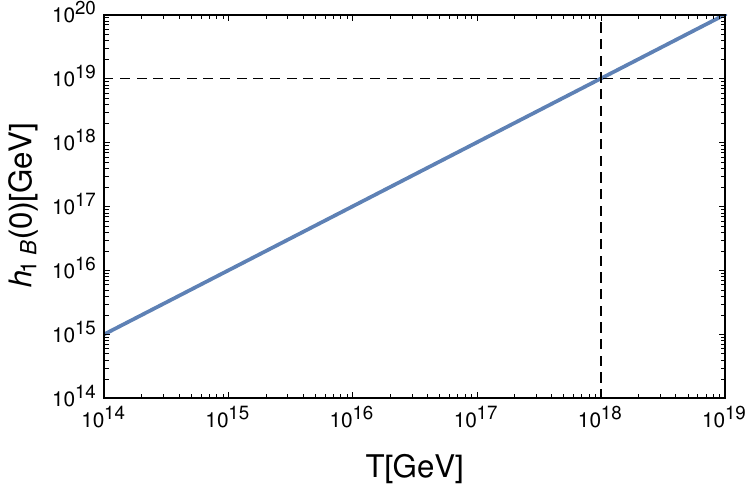}\label{f2a}}.}\caption{(a) variation of the effective potential $V/4V_{MAX}$ with $h_{1B}(0)/T$ for one particular temperature $T=10^{18.0}$ which leads to the center of bounce till Planck scale. The ratio $h_{1B}(0)/T$ always comes out to be $ \sim 14.75$ throughout the analysis which is shown by the trip of the arrow. (b) shows the variation of the center of bounce with different temperatures and the dashed black line denotes the center of bounce at Planck scale and the corresponding value of the temperature.  }\label{fig:phib0idm12}
	\end{center}
\end{figure}

After computing the profile of the bounce, the variation of the action with the temperature in GeV is plotted in \autoref{fig:action1}. Since, action is function of the field, more is the value of field profile, more is the value of corresponding action. In this scenario, the value of the bounce is achieved till Planck scale for temperature, $T=10^{18.0}$ GeV, which is shown by the tip of arrow in \autoref{fig:phib0idm12}(a). The variation of the bounce solution at the center with the temperature in GeV is given in \autoref{fig:phib0idm12}(b). The dashed black line again represents the field profile till Planck scale and the corresponding temperature.

\begin{figure}[H]
	\begin{center}
		\mbox{
			\subfigure[$\lambda_i = 0.01$, $ Y_N=0.01$ ]{\includegraphics[width=0.48\linewidth,angle=-0]{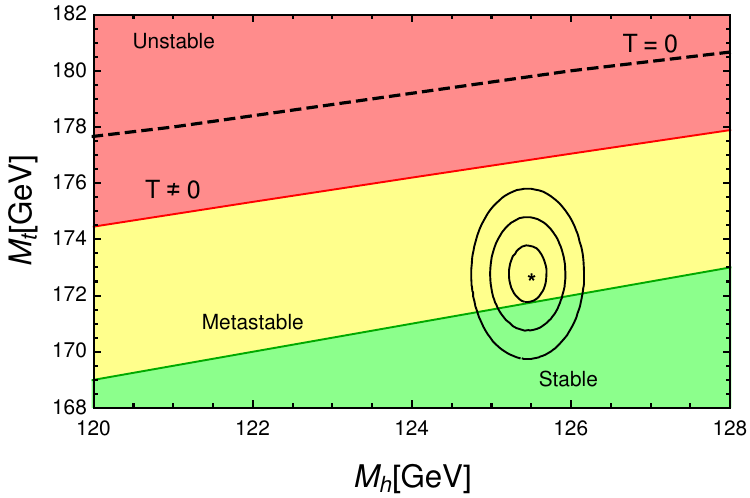}\label{f2a}}.}\caption{Phase space plot for the Higgs boson mass vs top quark mass. The green and the yellow region denotes the stable and the metastable region, respectively. The red line and the black dashed line corresponds to the instability at finite-temperature and the zero temperature.}\label{fig:diffprob1}
	\end{center}
\end{figure}

The phase space plot for the top-quark mass vs Higgs boson mass is depicted by \autoref{fig:diffprob1} for $m_t = 172.69 $ and $\lambda_i=0.01, \ Y_N=0.01$. The green and the yellow region again denotes the stable and the metastable region. The red line and the black dashed line corresponds to the unstable region at the finite temperature and the zero temperature. The lower value of the top-quark Yukawa coupling lead to the more stable region in comparison to the previous case and again, the instability bound is more strong at finite temperature as compared to the zero temperature bound.

\subsection{Case-III, $(m_t = 172.69- \delta m - 0.3), \lambda_i=0.01, \ Y_N=0.01$}

\begin{figure}[H]
	\begin{center}
		\mbox{
			\subfigure[$\lambda_i = 0.01$, $ Y_N=0.01$ ]{\includegraphics[width=0.48\linewidth,angle=-0]{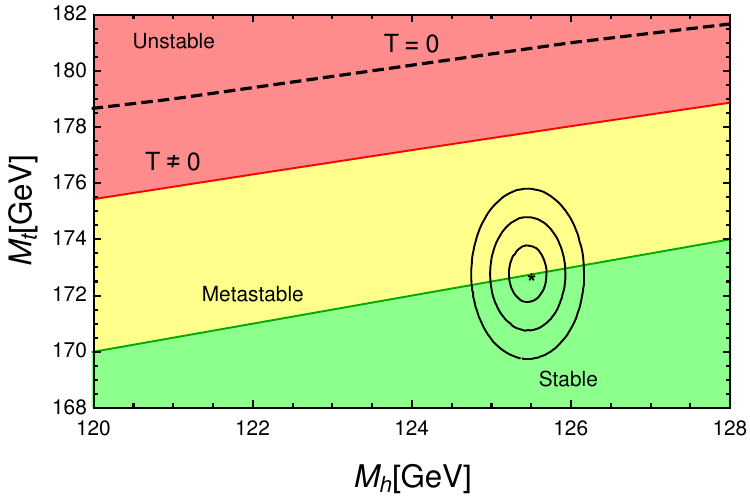}\label{f2a}}.}\caption{Phase space plot for the Higgs boson mass vs top-quark mass for $ m_t = 172.69- \delta m - 0.3 $. The green and the yellow region denotes the stable and the metastable region, respectively. The red line and the black dashed line corresponds to the instability at finite-temperature and the zero temperature.}\label{fig:phasespaceidm3}
	\end{center}
\end{figure}

In this scenario, the top-yukawa coupling is further reduced and hence, the corresponding negative contribution to the running of couplings is also reduced. Therefore, the stability bound is improved further and the instability bound is relaxed for both zero-temperature and the finite-temperature as shown in \autoref{fig:phasespaceidm3} for $m_t = 172.69- \delta m - 0.3 $ and $\lambda_i=0.01, \ Y_N=0.01$.

\begin{figure}[H]
	\begin{center}
		\mbox{\subfigure[$\lambda_i = 0.01$, $ Y_N=0.01$ ]{\includegraphics[width=0.48\linewidth,angle=-0]{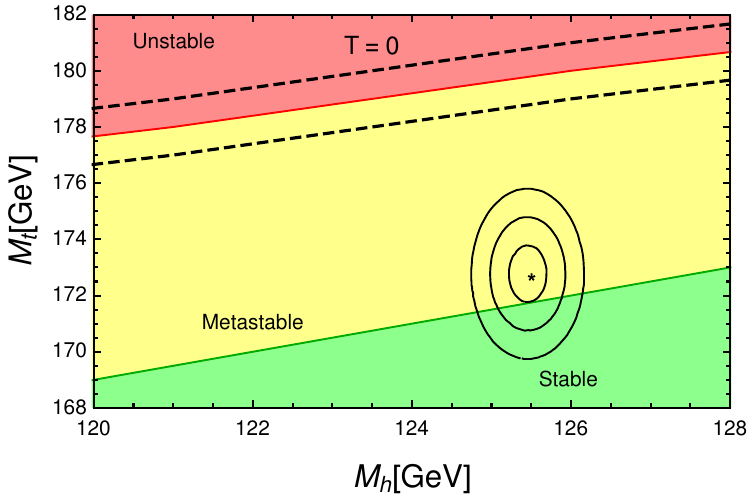}\label{f1a}}
			\subfigure[$\lambda_i = 0.01$, $ Y_N=0.01$ ]{\includegraphics[width=0.48\linewidth,angle=-0]{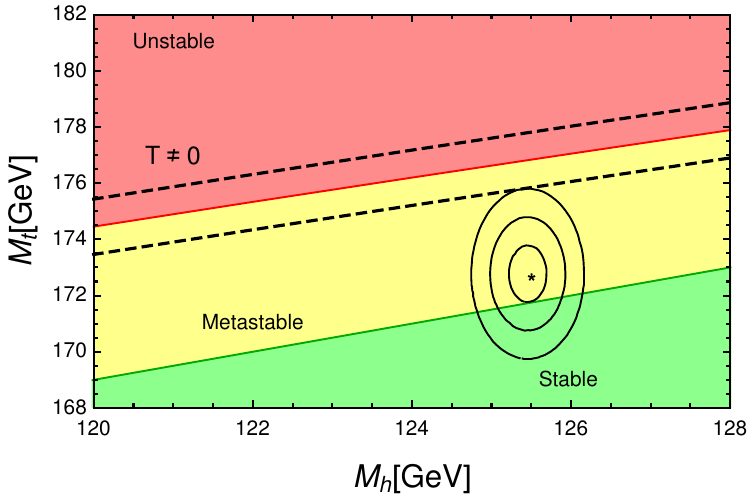}\label{f2a}}.}\caption{(a) Phase space plot for the Higgs boson mass vs top-quark mass at zero temperature, (b) Phase space plot for the Higgs mass vs top-quark mass at finite temperature. The green, yellow and the red regions corresponds to the unstable regions, respectively. The black dot at the cneter and the contours represents the central values for the Higgs mass and the top quark mass and the $1 \sigma, \ 2 \sigma $ and $3 \sigma$ uncertainities, respectively. The black dashed lines denotes the uncertainity in the top-quark mass in GeV.}\label{fig:phasespaceidm45}
	\end{center}
\end{figure}

The final phase space plot considering the uncertainity in the top-quark mass is given in \autoref{fig:phasespaceidm45} for $\lambda_i = 0.01$, $ Y_N=0.01$. The color code denotes the stable, metastable and the unstable regions similar to the previous scenarios. The black dashed lines denotes the boundary of the unstable region considering the top-quark mass uncertainity at zero temperature in \autoref{fig:phasespaceidm45}(a) and at finite-temperature in \autoref{fig:phasespaceidm45}(b). It is clear from \autoref{fig:phasespaceidm45}(a) to \autoref{fig:phasespaceidm45}(b) that the thermal fluctuations increases the chances for the tunneling from the false vacuua to the true vacuua and hence, the instability bound becomes more stringent. If the values of quartic couplings $\lambda_i's$ where $i \in 2,3,4,5$ are increased to 0.1 or even higher for same values of $Y_N =0.01$, the positive contribution increases a lot and the the $1 \sigma, \ 2 \sigma $ and $3 \sigma$ uncertainities lie completely in the stable region.

For further higher values of quartic couplings, i.e., $\lambda_i=0.1$ or above with $i \in {2,3,4,5}$, the effective quartic coupling remains positive all the way till Planck scale. Hence, the electroweak vacuum is stable for these higher values.

\section{Conclusion}
We have considered the extension of the SM with another $SU(2)$ Higgs doublet augmented with right-handed neutrinos. The doublet field acts as the inflaton field and the inflationary bounds from the spectral index and the tensor to scalar ratio are satisfied for the interaction quartic couplings $\lambda_3 \sim 1.0$ and $\lambda_4, \lambda_5 <1$. The reheating temperature is obtained as $10^{14}$ GeV. These couplings also comes out to be consistent with the electroweak phase transition (EWPT) dynamics. The allowed values from the GUT scale perturbative unitarity constraints the doublet mass parameter to be $m_{22}=400.0$ GeV from the strongly first-order phase transition for $Y_N=0.01$, and strongly first-order phase transition is no longer possible to satify for $Y_N=0.4$. The increase in the value of $Y_N$ reduces the strength of the phase transition by reducing the maximum allowed value of the interaction quartic couplings. 

For perturbative unitarity till Planck scale, the doublet mass parameter is restricted to 70 GeV from first-order phase transition for $Y_N=0.01$, and none of the value for the doublet mass parameter satisfies the first-order phase transition for $Y_N=0.4$. This actually puts a bound for the allowed value of the $Y_N$ from electroweak phase transition. Here also, the values allowed from the Planck scale perturbative unitarity are also consistent with the inflation bounds.\\

The stability of the electroweak vacuum both at the zero temperature and including the thermal corrections is also studied. The doublet field provides enough positive contribution to the running of couplings and the current experimentally measured values of the Higgs boson mass and the top-quark mass lie in the stable region until $Y_N=0.32$ at zero temperature. The thermal corrections trigger the tunneling probability from the false vacuum to the true vacuum and this makes the potential more unstable. The Higgs boson mass and the top-quark mass still lies in the stable region until the reheating temperature even after adding the thermal corrections. The $3 \sigma$ contour just hits the instability bound at finite-temperature considering the uncertainities in the top-quark mass for $\lambda_i =0.01$ and $Y_N=0.01$ with $i \in \{2,3,4,5\}$. The electroweak vacuum is more or less fully stable for further higher values of the quartic couplings, i.e., $\lambda_i =0.1$ and above with $i \in \{2,3,4,5\}$ \\

The second doublet is considered to be odd under the $Z_2$ symmetry and hence, it acts as the dark matter candidate after the inflation ends. The inert doublet freezes-out as a cold relic and thus becomes a cold dark matter candidate \cite{Choubey:2017hsq}. The bound on the DM mass satisfying the correct relic density bound comes out to be 1.38 TeV for $\lambda_3=0.5$. This much higher mass can only be consistent with the EWPT demanding the perturbative unitarity for scales lower than the GUT scale.

\section*{Acknowledgement}
This work is supported by the National Research Foundation of Korea (NRF) grant funded by the Korea government (MSIT) RS-2023-00283129 and RS-2024-00340153 (S.C.P.). This research was supported by an appointment to the YST Program at the APCTP through the Science and Technology Promotion Fund and Lottery Fund of the Korean Government. This was also supported by the Korean Local Governments - Gyeongsangbuk-do Province and Pohang city (S.J.). S.J. thanks Luigi Delle Rose for useful guidance in computing the finite-temperature stability of the EW vacuum and Denis Comelli for computation of thermal corrections for the bosons.

\appendix
\section{Two-loop $\beta$-functions for dimensionless couplings} \label{beta}
\subsection{Scalar Quartic Couplings}\label{A1}
\footnotesize{
	\begingroup
	\allowdisplaybreaks
	\begin{align*}
		\beta_{\lambda_{1}} \ =  \ &
		\frac{1}{16\pi^2} \Bigg[+\frac{27}{200} g_{1}^{4} +\frac{9}{20} g_{1}^{2} g_{2}^{2} +\frac{9}{8} g_{2}^{4} -\frac{9}{5} g_{1}^{2} \lambda_1 -9 g_{2}^{2} \lambda_1 +24 \lambda_{1}^{2} +2 \lambda_{3}^{2} +2 \lambda_3 \lambda_4 +\lambda_{4}^{2}+4 \lambda_{5}^{2}+12 \lambda_1 \mbox{Tr}\Big({Y_d  Y_{d}^{\dagger}}\Big) +4 \lambda_1 \mbox{Tr}\Big({Y_e  Y_{e}^{\dagger}}\Big) \nonumber \\ 
		& +4 \lambda_1 \mbox{Tr}\Big({Y_{N}  Y_{N}^{\dagger}}\Big) +12 \lambda_1 \mbox{Tr}\Big({Y_u  Y_{u}^{\dagger}}\Big) -6 \mbox{Tr}\Big({Y_d  Y_{d}^{\dagger}  Y_d  Y_{d}^{\dagger}}\Big) -2 \mbox{Tr}\Big({Y_e  Y_{e}^{\dagger}  Y_e  Y_{e}^{\dagger}}\Big) -2 \mbox{Tr}\Big({Y_{N}  Y_{N}^{\dagger}  Y_{N}  Y_{N}^{\dagger}}\Big) -6 \mbox{Tr}\Big({Y_u  Y_{u}^{\dagger}  Y_u  Y_{u}^{\dagger}}\Big) \Bigg] \nonumber \\
		& +\frac{1}{(16\pi^2)^2}\Bigg[-\frac{3537}{2000} g_{1}^{6} -\frac{1719}{400} g_{1}^{4} g_{2}^{2} -\frac{303}{80} g_{1}^{2} g_{2}^{4} +\frac{291}{16} g_{2}^{6} +\frac{1953}{200} g_{1}^{4} \lambda_1 +\frac{117}{20} g_{1}^{2} g_{2}^{2} \lambda_1 -\frac{51}{8} g_{2}^{4} \lambda_1 +\frac{108}{5} g_{1}^{2} \lambda_{1}^{2}+108 g_{2}^{2} \lambda_{1}^{2} \nonumber \\ 
		& -312 \lambda_{1}^{3} +\frac{9}{10} g_{1}^{4} \lambda_3 +\frac{15}{2} g_{2}^{4} \lambda_3 +\frac{12}{5} g_{1}^{2} \lambda_{3}^{2} +12 g_{2}^{2} \lambda_{3}^{2} -20 \lambda_1 \lambda_{3}^{2} -8 \lambda_{3}^{3} +\frac{9}{20} g_{1}^{4} \lambda_4 +\frac{3}{2} g_{1}^{2} g_{2}^{2} \lambda_4 +\frac{15}{4} g_{2}^{4} \lambda_4 +\frac{12}{5} g_{1}^{2} \lambda_3 \lambda_4 \nonumber \\ 
		& +12 g_{2}^{2} \lambda_3 \lambda_4 -20 \lambda_1 \lambda_3 \lambda_4 -12 \lambda_{3}^{2} \lambda_4 +\frac{6}{5} g_{1}^{2} \lambda_{4}^{2} +3 g_{2}^{2} \lambda_{4}^{2} -12 \lambda_1 \lambda_{4}^{2} -16 \lambda_3 \lambda_{4}^{2} -6 \lambda_{4}^{3} -\frac{12}{5} g_{1}^{2} \lambda_{5}^{2} -56 \lambda_1 \lambda_{5}^{2} -80 \lambda_3 \lambda_{5}^{2} -88 \lambda_4 \lambda_{5}^{2} \nonumber \\ 
		&+\frac{1}{20} \Big(-5 \Big(64 \lambda_1 \Big(-5 g_{3}^{2}  + 9 \lambda_1 \Big) -90 g_{2}^{2} \lambda_1  + 9 g_{2}^{4} \Big) + 9 g_{1}^{4}  + g_{1}^{2} \Big(50 \lambda_1  + 54 g_{2}^{2} \Big)\Big)\mbox{Tr}\Big({Y_d  Y_{d}^{\dagger}}\Big)-\frac{3}{20} \Big(15 g_{1}^{4}  -2 g_{1}^{2} \Big(11 g_{2}^{2}  + 25 \lambda_1 \Big) \nonumber \\ 
		& + 5 \Big(-10 g_{2}^{2} \lambda_1  + 64 \lambda_{1}^{2}  + g_{2}^{4}\Big)\Big)\mbox{Tr}\Big({Y_e  Y_{e}^{\dagger}}\Big) -\frac{9}{100} g_{1}^{4} \mbox{Tr}\Big({Y_{N}  Y_{N}^{\dagger}}\Big)-\frac{3}{10} g_{1}^{2} g_{2}^{2} \mbox{Tr}\Big({Y_{N}  Y_{N}^{\dagger}}\Big) -\frac{3}{4} g_{2}^{4} \mbox{Tr}\Big({Y_{N}  Y_{N}^{\dagger}}\Big) +\frac{3}{2} g_{1}^{2} \lambda_1 \mbox{Tr}\Big({Y_{N}  Y_{N}^{\dagger}}\Big)  \nonumber \\ 
		&+\frac{15}{2} g_{2}^{2} \lambda_1 \mbox{Tr}\Big({Y_{N}  Y_{N}^{\dagger}}\Big)-48 \lambda_{1}^{2} \mbox{Tr}\Big({Y_{N}  Y_{N}^{\dagger}}\Big) -\frac{171}{100} g_{1}^{4} \mbox{Tr}\Big({Y_u  Y_{u}^{\dagger}}\Big) +\frac{63}{10} g_{1}^{2} g_{2}^{2} \mbox{Tr}\Big({Y_u  Y_{u}^{\dagger}}\Big) -\frac{9}{4} g_{2}^{4} \mbox{Tr}\Big({Y_u  Y_{u}^{\dagger}}\Big)+\frac{17}{2} g_{1}^{2} \lambda_1 \mbox{Tr}\Big({Y_u  Y_{u}^{\dagger}}\Big) \nonumber \\ 
		& +\frac{45}{2} g_{2}^{2} \lambda_1 \mbox{Tr}\Big({Y_u  Y_{u}^{\dagger}}\Big) +80 g_{3}^{2} \lambda_1 \mbox{Tr}\Big({Y_u  Y_{u}^{\dagger}}\Big) -144 \lambda_{1}^{2} \mbox{Tr}\Big({Y_u  Y_{u}^{\dagger}}\Big)+\frac{4}{5} g_{1}^{2} \mbox{Tr}\Big({Y_d  Y_{d}^{\dagger}  Y_d  Y_{d}^{\dagger}}\Big) -32 g_{3}^{2} \mbox{Tr}\Big({Y_d  Y_{d}^{\dagger}  Y_d  Y_{d}^{\dagger}}\Big) -3 \lambda_1 \mbox{Tr}\Big({Y_d  Y_{d}^{\dagger}  Y_d  Y_{d}^{\dagger}}\Big) \nonumber \\ 
		& -42 \lambda_1 \mbox{Tr}\Big({Y_d  Y_{d}^{\dagger}  Y_u  Y_{u}^{\dagger}}\Big) -\frac{12}{5} g_{1}^{2} \mbox{Tr}\Big({Y_e  Y_{e}^{\dagger}  Y_e  Y_{e}^{\dagger}}\Big) - \lambda_1 \mbox{Tr}\Big({Y_e  Y_{e}^{\dagger}  Y_e  Y_{e}^{\dagger}}\Big) -14 \lambda_1 \mbox{Tr}\Big({Y_e  Y_{e}^{\dagger}  Y_{N}^{T}  Y_{N}^*}\Big) - \lambda_1 \mbox{Tr}\Big({Y_{N}  Y_{N}^{\dagger}  Y_{N}  Y_{N}^{\dagger}}\Big) \nonumber \\ 
		&-\frac{8}{5} g_{1}^{2} \mbox{Tr}\Big({Y_u  Y_{u}^{\dagger}  Y_u  Y_{u}^{\dagger}}\Big) -32 g_{3}^{2} \mbox{Tr}\Big({Y_u  Y_{u}^{\dagger}  Y_u  Y_{u}^{\dagger}}\Big) -3 \lambda_1 \mbox{Tr}\Big({Y_u  Y_{u}^{\dagger}  Y_u  Y_{u}^{\dagger}}\Big) +30 \mbox{Tr}\Big({Y_d  Y_{d}^{\dagger}  Y_d  Y_{d}^{\dagger}  Y_d  Y_{d}^{\dagger}}\Big) -6 \mbox{Tr}\Big({Y_d  Y_{d}^{\dagger}  Y_u  Y_{u}^{\dagger}  Y_d  Y_{d}^{\dagger}}\Big)\nonumber \\ 
		& -6 \mbox{Tr}\Big({Y_d  Y_{d}^{\dagger}  Y_u  Y_{u}^{\dagger}  Y_u  Y_{u}^{\dagger}}\Big) +10 \mbox{Tr}\Big({Y_e  Y_{e}^{\dagger}  Y_e  Y_{e}^{\dagger}  Y_e  Y_{e}^{\dagger}}\Big) -2 \mbox{Tr}\Big({Y_e  Y_{e}^{\dagger}  Y_e  Y_{e}^{\dagger}  Y_{N}^{T}  Y_{N}^*}\Big) -2 \mbox{Tr}\Big({Y_e  Y_{e}^{\dagger}  Y_{N}^{T}  Y_{N}^*  Y_{N}^{T}  Y_{N}^*}\Big)\nonumber \\ 
		& +10 \mbox{Tr}\Big({Y_{N}  Y_{N}^{\dagger}  Y_{N}  Y_{N}^{\dagger}  Y_{N}  Y_{N}^{\dagger}}\Big) +30 \mbox{Tr}\Big({Y_u  Y_{u}^{\dagger}  Y_u  Y_{u}^{\dagger}  Y_u  Y_{u}^{\dagger}}\Big)\Bigg] \, . \\
		\beta_{\lambda_2} \  = \ &
		\frac{1}{16\pi^2}\Bigg[
		24 \lambda_{2}^{2}  + 2 \lambda_{3}^{2}  + 2 \lambda_3 \lambda_4  + 4 \lambda_{5}^{2}  -9 g_{2}^{2} \lambda_2  + \frac{27}{200} g_{1}^{4}  + \frac{9}{20} g_{1}^{2} \Big(-4 \lambda_2  + g_{2}^{2}\Big) + \frac{9}{8} g_{2}^{4}  + \lambda_{4}^{2}\Bigg] \nonumber \\
		& +\frac{1}{(16\pi^2)^2}\Bigg[-\frac{3537}{2000} g_{1}^{6} -\frac{1719}{400} g_{1}^{4} g_{2}^{2} -\frac{303}{80} g_{1}^{2} g_{2}^{4} +\frac{291}{16} g_{2}^{6} +\frac{1953}{200} g_{1}^{4} \lambda_2 +\frac{117}{20} g_{1}^{2} g_{2}^{2} \lambda_2 -\frac{51}{8} g_{2}^{4} \lambda_2 +\frac{108}{5} g_{1}^{2} \lambda_{2}^{2} +108 g_{2}^{2} \lambda_{2}^{2}\nonumber \\ 
		& -312 \lambda_{2}^{3} +\frac{9}{10} g_{1}^{4} \lambda_3 +\frac{15}{2} g_{2}^{4} \lambda_3 +\frac{12}{5} g_{1}^{2} \lambda_{3}^{2} +12 g_{2}^{2} \lambda_{3}^{2} -20 \lambda_2 \lambda_{3}^{2} -8 \lambda_{3}^{3} +\frac{9}{20} g_{1}^{4} \lambda_4 +\frac{3}{2} g_{1}^{2} g_{2}^{2} \lambda_4 +\frac{15}{4} g_{2}^{4} \lambda_4 +\frac{12}{5} g_{1}^{2} \lambda_3 \lambda_4+12 g_{2}^{2} \lambda_3 \lambda_4 \nonumber \\ 
		&  -20 \lambda_2 \lambda_3 \lambda_4 -12 \lambda_{3}^{2} \lambda_4 +\frac{6}{5} g_{1}^{2} \lambda_{4}^{2} +3 g_{2}^{2} \lambda_{4}^{2} -12 \lambda_2 \lambda_{4}^{2} -16 \lambda_3 \lambda_{4}^{2} -6 \lambda_{4}^{3} -\frac{12}{5} g_{1}^{2} \lambda_{5}^{2} -56 \lambda_2 \lambda_{5}^{2} -80 \lambda_3 \lambda_{5}^{2} -88 \lambda_4 \lambda_{5}^{2}-4 \lambda_{3}^{2} \mbox{Tr}\Big({Y_{N}  Y_{N}^{\dagger}}\Big) \nonumber \\ 
		&-6 \Big(2 \lambda_{3}^{2}  + 2 \lambda_3 \lambda_4  + 4 \lambda_{5}^{2}  + \lambda_{4}^{2}\Big)\mbox{Tr}\Big({Y_d  Y_{d}^{\dagger}}\Big) -2 \Big(2 \lambda_{3}^{2}  + 2 \lambda_3 \lambda_4  + 4 \lambda_{5}^{2}  + \lambda_{4}^{2}\Big)\mbox{Tr}\Big({Y_e  Y_{e}^{\dagger}}\Big)-4 \lambda_3 \lambda_4 \mbox{Tr}\Big({Y_{N}  Y_{N}^{\dagger}}\Big) -2 \lambda_{4}^{2} \mbox{Tr}\Big({Y_{N}  Y_{N}^{\dagger}}\Big) \nonumber \\ 
		&  -8 \lambda_{5}^{2} \mbox{Tr}\Big({Y_{N}  Y_{N}^{\dagger}}\Big) -12 \lambda_{3}^{2} \mbox{Tr}\Big({Y_u  Y_{u}^{\dagger}}\Big) -12 \lambda_3 \lambda_4 \mbox{Tr}\Big({Y_u  Y_{u}^{\dagger}}\Big) -6 \lambda_{4}^{2} \mbox{Tr}\Big({Y_u  Y_{u}^{\dagger}}\Big) -24 \lambda_{5}^{2} \mbox{Tr}\Big({Y_u  Y_{u}^{\dagger}}\Big) \Bigg] \, .  \\
		\beta_{\lambda_{3}} \ =  \ &
		\frac{1}{16\pi^2}\Bigg[+\frac{27}{100} g_{1}^{4} -\frac{9}{10} g_{1}^{2} g_{2}^{2} +\frac{9}{4} g_{2}^{4} -\frac{9}{5} g_{1}^{2} \lambda_3 -9 g_{2}^{2} \lambda_3 +12 \lambda_1 \lambda_3 +12 \lambda_2 \lambda_3 +4 \lambda_{3}^{2} +4 \lambda_1 \lambda_4 +4 \lambda_2 \lambda_4 +2 \lambda_{4}^{2}+8 \lambda_{5}^{2} +6 \lambda_3 \mbox{Tr}\Big({Y_d  Y_{d}^{\dagger}}\Big) \nonumber \\ 
		& +2 \lambda_3 \mbox{Tr}\Big({Y_e  Y_{e}^{\dagger}}\Big) +2 \lambda_3 \mbox{Tr}\Big({Y_{N}  Y_{N}^{\dagger}}\Big) +6 \lambda_3 \mbox{Tr}\Big({Y_u  Y_{u}^{\dagger}}\Big)\Bigg]\nonumber \\
		&+\frac{1}{(16\pi^2)^2}\Bigg[-\frac{3537}{1000} g_{1}^{6} +\frac{909}{200} g_{1}^{4} g_{2}^{2} +\frac{33}{40} g_{1}^{2} g_{2}^{4} +\frac{291}{8} g_{2}^{6} +\frac{27}{10} g_{1}^{4} \lambda_1 -3 g_{1}^{2} g_{2}^{2} \lambda_1 +\frac{45}{2} g_{2}^{4} \lambda_1 +\frac{27}{10} g_{1}^{4} \lambda_2 -3 g_{1}^{2} g_{2}^{2} \lambda_2 +\frac{45}{2} g_{2}^{4} \lambda_2\nonumber \\ 
		& +\frac{1773}{200} g_{1}^{4} \lambda_3 +\frac{33}{20} g_{1}^{2} g_{2}^{2} \lambda_3 -\frac{111}{8} g_{2}^{4} \lambda_3 +\frac{72}{5} g_{1}^{2} \lambda_1 \lambda_3 +72 g_{2}^{2} \lambda_1 \lambda_3 -60 \lambda_{1}^{2} \lambda_3 +\frac{72}{5} g_{1}^{2} \lambda_2 \lambda_3 +72 g_{2}^{2} \lambda_2 \lambda_3 -60 \lambda_{2}^{2} \lambda_3 +\frac{6}{5} g_{1}^{2} \lambda_{3}^{2}\nonumber \\ 
		&  +6 g_{2}^{2} \lambda_{3}^{2} -72 \lambda_1 \lambda_{3}^{2} -72 \lambda_2 \lambda_{3}^{2} -12 \lambda_{3}^{3} +\frac{9}{10} g_{1}^{4} \lambda_4 -\frac{9}{5} g_{1}^{2} g_{2}^{2} \lambda_4 +\frac{15}{2} g_{2}^{4} \lambda_4 +\frac{24}{5} g_{1}^{2} \lambda_1 \lambda_4 +36 g_{2}^{2} \lambda_1 \lambda_4 -16 \lambda_{1}^{2} \lambda_4 +\frac{24}{5} g_{1}^{2} \lambda_2 \lambda_4 \nonumber \\ 
		&+36 g_{2}^{2} \lambda_2 \lambda_4 -16 \lambda_{2}^{2} \lambda_4 -12 g_{2}^{2} \lambda_3 \lambda_4 -32 \lambda_1 \lambda_3 \lambda_4 -32 \lambda_2 \lambda_3 \lambda_4 -4 \lambda_{3}^{2} \lambda_4 -\frac{6}{5} g_{1}^{2} \lambda_{4}^{2}+6 g_{2}^{2} \lambda_{4}^{2} -28 \lambda_1 \lambda_{4}^{2} -28 \lambda_2 \lambda_{4}^{2} -16 \lambda_3 \lambda_{4}^{2} -12 \lambda_{4}^{3} \nonumber \\ 
		& +\frac{48}{5} g_{1}^{2} \lambda_{5}^{2} -144 \lambda_1 \lambda_{5}^{2} -144 \lambda_2 \lambda_{5}^{2}-72 \lambda_3 \lambda_{5}^{2} -176 \lambda_4 \lambda_{5}^{2} -\frac{9}{100} g_{1}^{4} \mbox{Tr}\Big({Y_{N}  Y_{N}^{\dagger}}\Big) +\frac{3}{10} g_{1}^{2} g_{2}^{2} \mbox{Tr}\Big({Y_{N}  Y_{N}^{\dagger}}\Big) -\frac{3}{4} g_{2}^{4} \mbox{Tr}\Big({Y_{N}  Y_{N}^{\dagger}}\Big)  \nonumber \\ 
		&+\frac{1}{20} \Big(-5 \Big(-45 g_{2}^{2} \lambda_3  + 8 \Big(-20 g_{3}^{2} \lambda_3  + 3 \Big(2 \lambda_{3}^{2}  + 4 \lambda_1 \Big(3 \lambda_3  + \lambda_4\Big) + 4 \lambda_{5}^{2}  + \lambda_{4}^{2}\Big)\Big) + 9 g_{2}^{4} \Big) + 9 g_{1}^{4}  + g_{1}^{2} \Big(25 \lambda_3  -54 g_{2}^{2} \Big)\Big)\mbox{Tr}\Big({Y_d  Y_{d}^{\dagger}}\Big) \nonumber \\ 
		&-\frac{1}{20} \Big(45 g_{1}^{4}  + 5 \Big(-15 g_{2}^{2} \lambda_3  + 3 g_{2}^{4}  + 8 \Big(2 \lambda_{3}^{2}  + 4 \lambda_1 \Big(3 \lambda_3  + \lambda_4\Big) + 4 \lambda_{5}^{2}  + \lambda_{4}^{2}\Big)\Big) + g_{1}^{2} \Big(66 g_{2}^{2}  -75 \lambda_3 \Big)\Big)\mbox{Tr}\Big({Y_e  Y_{e}^{\dagger}}\Big) +\frac{3}{4} g_{1}^{2} \lambda_3 \mbox{Tr}\Big({Y_{N}  Y_{N}^{\dagger}}\Big)\nonumber \\ 
		&+\frac{15}{4} g_{2}^{2} \lambda_3 \mbox{Tr}\Big({Y_{N}  Y_{N}^{\dagger}}\Big) -24 \lambda_1 \lambda_3 \mbox{Tr}\Big({Y_{N}  Y_{N}^{\dagger}}\Big) -4 \lambda_{3}^{2} \mbox{Tr}\Big({Y_{N}  Y_{N}^{\dagger}}\Big) -8 \lambda_1 \lambda_4 \mbox{Tr}\Big({Y_{N}  Y_{N}^{\dagger}}\Big) -2 \lambda_{4}^{2} \mbox{Tr}\Big({Y_{N}  Y_{N}^{\dagger}}\Big) -8 \lambda_{5}^{2} \mbox{Tr}\Big({Y_{N}  Y_{N}^{\dagger}}\Big)\nonumber \\ 
		& -\frac{171}{100} g_{1}^{4} \mbox{Tr}\Big({Y_u  Y_{u}^{\dagger}}\Big) -\frac{63}{10} g_{1}^{2} g_{2}^{2} \mbox{Tr}\Big({Y_u  Y_{u}^{\dagger}}\Big) -\frac{9}{4} g_{2}^{4} \mbox{Tr}\Big({Y_u  Y_{u}^{\dagger}}\Big) +\frac{17}{4} g_{1}^{2} \lambda_3 \mbox{Tr}\Big({Y_u  Y_{u}^{\dagger}}\Big) +\frac{45}{4} g_{2}^{2} \lambda_3 \mbox{Tr}\Big({Y_u  Y_{u}^{\dagger}}\Big) +40 g_{3}^{2} \lambda_3 \mbox{Tr}\Big({Y_u  Y_{u}^{\dagger}}\Big) \nonumber \\ 
		&-72 \lambda_1 \lambda_3 \mbox{Tr}\Big({Y_u  Y_{u}^{\dagger}}\Big) -12 \lambda_{3}^{2} \mbox{Tr}\Big({Y_u  Y_{u}^{\dagger}}\Big) -24 \lambda_1 \lambda_4 \mbox{Tr}\Big({Y_u  Y_{u}^{\dagger}}\Big) -6 \lambda_{4}^{2} \mbox{Tr}\Big({Y_u  Y_{u}^{\dagger}}\Big)-24 \lambda_{5}^{2} \mbox{Tr}\Big({Y_u  Y_{u}^{\dagger}}\Big) -\frac{27}{2} \lambda_3 \mbox{Tr}\Big({Y_d  Y_{d}^{\dagger}  Y_d  Y_{d}^{\dagger}}\Big) \nonumber \\ 
		& -21 \lambda_3 \mbox{Tr}\Big({Y_d  Y_{d}^{\dagger}  Y_u  Y_{u}^{\dagger}}\Big) -24 \lambda_4 \mbox{Tr}\Big({Y_d  Y_{d}^{\dagger}  Y_u  Y_{u}^{\dagger}}\Big) -\frac{9}{2} \lambda_3 \mbox{Tr}\Big({Y_e  Y_{e}^{\dagger}  Y_e  Y_{e}^{\dagger}}\Big) -7 \lambda_3 \mbox{Tr}\Big({Y_e  Y_{e}^{\dagger}  Y_{N}^{T}  Y_{N}^*}\Big) -8 \lambda_4 \mbox{Tr}\Big({Y_e  Y_{e}^{\dagger}  Y_{N}^{T}  Y_{N}^*}\Big)  \nonumber \\ 
		&-\frac{9}{2} \lambda_3 \mbox{Tr}\Big({Y_{N}  Y_{N}^{\dagger}  Y_{N}  Y_{N}^{\dagger}}\Big)-\frac{27}{2} \lambda_3 \mbox{Tr}\Big({Y_u  Y_{u}^{\dagger}  Y_u  Y_{u}^{\dagger}}\Big) 
		 \Bigg] \, . \\
		\beta_{\lambda_4} \  = \ &  
		\frac{1}{16\pi^2}\Bigg[+\frac{9}{5} g_{1}^{2} g_{2}^{2} -\frac{9}{5} g_{1}^{2} \lambda_4 -9 g_{2}^{2} \lambda_4 +4 \lambda_1 \lambda_4 +4 \lambda_2 \lambda_4 +8 \lambda_3 \lambda_4 +4 \lambda_{4}^{2} +32 \lambda_{5}^{2} +6 \lambda_4 \mbox{Tr}\Big({Y_d  Y_{d}^{\dagger}}\Big) +2 \lambda_4 \mbox{Tr}\Big({Y_e  Y_{e}^{\dagger}}\Big) +2 \lambda_4 \mbox{Tr}\Big({Y_{N}  Y_{N}^{\dagger}}\Big)\nonumber \\ 
		& +6 \lambda_4 \mbox{Tr}\Big({Y_u  Y_{u}^{\dagger}}\Big)\Bigg]\nonumber \\
		&+\frac{1}{(16\pi^2)^2}\Bigg[-\frac{657}{50} g_{1}^{4} g_{2}^{2} -\frac{42}{5} g_{1}^{2} g_{2}^{4} +6 g_{1}^{2} g_{2}^{2} \lambda_1 +6 g_{1}^{2} g_{2}^{2} \lambda_2 +\frac{6}{5} g_{1}^{2} g_{2}^{2} \lambda_3 +\frac{1413}{200} g_{1}^{4} \lambda_4 +\frac{153}{20} g_{1}^{2} g_{2}^{2} \lambda_4 -\frac{231}{8} g_{2}^{4} \lambda_4 +\frac{24}{5} g_{1}^{2} \lambda_1 \lambda_4\nonumber \\ 
		& -28 \lambda_{1}^{2} \lambda_4 +\frac{24}{5} g_{1}^{2} \lambda_2 \lambda_4 -28 \lambda_{2}^{2} \lambda_4 +\frac{12}{5} g_{1}^{2} \lambda_3 \lambda_4 +36 g_{2}^{2} \lambda_3 \lambda_4 -80 \lambda_1 \lambda_3 \lambda_4 -80 \lambda_2 \lambda_3 \lambda_4 -28 \lambda_{3}^{2} \lambda_4 +\frac{24}{5} g_{1}^{2} \lambda_{4}^{2}+18 g_{2}^{2} \lambda_{4}^{2} -40 \lambda_1 \lambda_{4}^{2} \nonumber \\ 
		& -40 \lambda_2 \lambda_{4}^{2} -28 \lambda_3 \lambda_{4}^{2}+\frac{192}{5} g_{1}^{2} \lambda_{5}^{2} +216 g_{2}^{2} \lambda_{5}^{2} -192 \lambda_1 \lambda_{5}^{2} -192 \lambda_2 \lambda_{5}^{2} -192 \lambda_3 \lambda_{5}^{2} -104 \lambda_4 \lambda_{5}^{2} + g_{1}^{2} \Big(\frac{27}{5} g_{2}^{2}  + \frac{5}{4} \lambda_4 \Big)\Big)\mbox{Tr}\Big({Y_d  Y_{d}^{\dagger}}\Big) \nonumber \\ 
		&+\Big(4 \Big(10 g_{3}^{2} \lambda_4  -3 \Big(2 \lambda_1 \lambda_4  + 2 \lambda_3 \lambda_4  + 8 \lambda_{5}^{2}  + \lambda_{4}^{2}\Big)\Big) + \frac{45}{4} g_{2}^{2} \lambda_4 -8 \lambda_3 \lambda_4 \mbox{Tr}\Big({Y_{N}  Y_{N}^{\dagger}}\Big) -4 \lambda_{4}^{2} \mbox{Tr}\Big({Y_{N}  Y_{N}^{\dagger}}\Big) -32 \lambda_{5}^{2} \mbox{Tr}\Big({Y_{N}  Y_{N}^{\dagger}}\Big)  \nonumber \\ 
		&+\frac{63}{5} g_{1}^{2} g_{2}^{2} \mbox{Tr}\Big({Y_u  Y_{u}^{\dagger}}\Big)+\Big(-4 \Big(2 \lambda_1 \lambda_4  + 2 \lambda_3 \lambda_4  + 8 \lambda_{5}^{2}  + \lambda_{4}^{2}\Big) + \frac{15}{4} g_{2}^{2} \lambda_4  + \frac{3}{20} g_{1}^{2} \Big(25 \lambda_4  + 44 g_{2}^{2} \Big)\Big)\mbox{Tr}\Big({Y_e  Y_{e}^{\dagger}}\Big)-\frac{3}{5} g_{1}^{2} g_{2}^{2} \mbox{Tr}\Big({Y_{N}  Y_{N}^{\dagger}}\Big) \nonumber \\ 
		& +\frac{3}{4} g_{1}^{2} \lambda_4 \mbox{Tr}\Big({Y_{N}  Y_{N}^{\dagger}}\Big) +\frac{15}{4} g_{2}^{2} \lambda_4 \mbox{Tr}\Big({Y_{N}  Y_{N}^{\dagger}}\Big) -8 \lambda_1 \lambda_4 \mbox{Tr}\Big({Y_{N}  Y_{N}^{\dagger}}\Big)+\frac{17}{4} g_{1}^{2} \lambda_4 \mbox{Tr}\Big({Y_u  Y_{u}^{\dagger}}\Big) +\frac{45}{4} g_{2}^{2} \lambda_4 \mbox{Tr}\Big({Y_u  Y_{u}^{\dagger}}\Big) +40 g_{3}^{2} \lambda_4 \mbox{Tr}\Big({Y_u  Y_{u}^{\dagger}}\Big) \nonumber \\  
		& -24 \lambda_1 \lambda_4 \mbox{Tr}\Big({Y_u  Y_{u}^{\dagger}}\Big)-24 \lambda_3 \lambda_4 \mbox{Tr}\Big({Y_u  Y_{u}^{\dagger}}\Big) -12 \lambda_{4}^{2} \mbox{Tr}\Big({Y_u  Y_{u}^{\dagger}}\Big) -96 \lambda_{5}^{2} \mbox{Tr}\Big({Y_u  Y_{u}^{\dagger}}\Big) -\frac{27}{2} \lambda_4 \mbox{Tr}\Big({Y_d  Y_{d}^{\dagger}  Y_d  Y_{d}^{\dagger}}\Big) \nonumber \\ 
		&+27 \lambda_4 \mbox{Tr}\Big({Y_d  Y_{d}^{\dagger}  Y_u  Y_{u}^{\dagger}}\Big) -\frac{9}{2} \lambda_4 \mbox{Tr}\Big({Y_e  Y_{e}^{\dagger}  Y_e  Y_{e}^{\dagger}}\Big) +9 \lambda_4 \mbox{Tr}\Big({Y_e  Y_{e}^{\dagger}  Y_{N}^{T}  Y_{N}^*}\Big) -\frac{9}{2} \lambda_4 \mbox{Tr}\Big({Y_{N}  Y_{N}^{\dagger}  Y_{N}  Y_{N}^{\dagger}}\Big) -\frac{27}{2} \lambda_4 \mbox{Tr}\Big({Y_u  Y_{u}^{\dagger}  Y_u  Y_{u}^{\dagger}}\Big)\Bigg] \, .  \\
		\beta_{\lambda_5} \  = \ &  
		\frac{1}{16\pi^2}\Bigg[-\frac{9}{5} g_{1}^{2} \lambda_5 -9 g_{2}^{2} \lambda_5 +4 \lambda_1 \lambda_5 +4 \lambda_2 \lambda_5 +8 \lambda_3 \lambda_5 +12 \lambda_4 \lambda_5 +6 \lambda_5 \mbox{Tr}\Big({Y_d  Y_{d}^{\dagger}}\Big) +2 \lambda_5 \mbox{Tr}\Big({Y_e  Y_{e}^{\dagger}}\Big) +2 \lambda_5 \mbox{Tr}\Big({Y_{N}  Y_{N}^{\dagger}}\Big)\nonumber \\ 
		& +6 \lambda_5 \mbox{Tr}\Big({Y_u  Y_{u}^{\dagger}}\Big) \Bigg]\nonumber \\
		&+\frac{1}{(16\pi^2)^2}\Bigg[ +\frac{1413}{200} g_{1}^{4} \lambda_5 +\frac{57}{20} g_{1}^{2} g_{2}^{2} \lambda_5 -\frac{231}{8} g_{2}^{4} \lambda_5 -\frac{12}{5} g_{1}^{2} \lambda_1 \lambda_5 -28 \lambda_{1}^{2} \lambda_5 -\frac{12}{5} g_{1}^{2} \lambda_2 \lambda_5 -28 \lambda_{2}^{2} \lambda_5 +\frac{48}{5} g_{1}^{2} \lambda_3 \lambda_5 +36 g_{2}^{2} \lambda_3 \lambda_5\nonumber \\ 
		& -80 \lambda_1 \lambda_3 \lambda_5 -80 \lambda_2 \lambda_3 \lambda_5 -28 \lambda_{3}^{2} \lambda_5 +\frac{72}{5} g_{1}^{2} \lambda_4 \lambda_5 +72 g_{2}^{2} \lambda_4 \lambda_5 -88 \lambda_1 \lambda_4 \lambda_5 -88 \lambda_2 \lambda_4 \lambda_5 -76 \lambda_3 \lambda_4 \lambda_5 -32 \lambda_{4}^{2} \lambda_5 +24 \lambda_{5}^{3} \nonumber \\ 
		&+\frac{1}{4} \Big(16 \Big(10 g_{3}^{2}  -6 \lambda_1  -6 \lambda_3  -9 \lambda_4 \Big) + 45 g_{2}^{2}  + 5 g_{1}^{2} \Big)\lambda_5 \mbox{Tr}\Big({Y_d  Y_{d}^{\dagger}}\Big)+\frac{1}{4} \Big(15 g_{1}^{2}  + 15 g_{2}^{2}  -16 \Big(2 \lambda_1  + 2 \lambda_3  + 3 \lambda_4 \Big)\Big)\lambda_5 \mbox{Tr}\Big({Y_e  Y_{e}^{\dagger}}\Big)  \nonumber \\ 
		&+\frac{3}{4} g_{1}^{2} \lambda_5 \mbox{Tr}\Big({Y_{N}  Y_{N}^{\dagger}}\Big) +\frac{15}{4} g_{2}^{2} \lambda_5 \mbox{Tr}\Big({Y_{N}  Y_{N}^{\dagger}}\Big) -8 \lambda_1 \lambda_5 \mbox{Tr}\Big({Y_{N}  Y_{N}^{\dagger}}\Big) -8 \lambda_3 \lambda_5 \mbox{Tr}\Big({Y_{N}  Y_{N}^{\dagger}}\Big) -12 \lambda_4 \lambda_5 \mbox{Tr}\Big({Y_{N}  Y_{N}^{\dagger}}\Big) +\frac{17}{4} g_{1}^{2} \lambda_5 \mbox{Tr}\Big({Y_u  Y_{u}^{\dagger}}\Big) \nonumber \\ 
		&+\frac{45}{4} g_{2}^{2} \lambda_5 \mbox{Tr}\Big({Y_u  Y_{u}^{\dagger}}\Big) +40 g_{3}^{2} \lambda_5 \mbox{Tr}\Big({Y_u  Y_{u}^{\dagger}}\Big) -24 \lambda_1 \lambda_5 \mbox{Tr}\Big({Y_u  Y_{u}^{\dagger}}\Big) -24 \lambda_3 \lambda_5 \mbox{Tr}\Big({Y_u  Y_{u}^{\dagger}}\Big)-36 \lambda_4 \lambda_5 \mbox{Tr}\Big({Y_u  Y_{u}^{\dagger}}\Big) -\frac{3}{2} \lambda_5 \mbox{Tr}\Big({Y_d  Y_{d}^{\dagger}  Y_d  Y_{d}^{\dagger}}\Big) \nonumber \\ 
		& +3 \lambda_5 \mbox{Tr}\Big({Y_d  Y_{d}^{\dagger}  Y_u  Y_{u}^{\dagger}}\Big) -\frac{1}{2} \lambda_5 \mbox{Tr}\Big({Y_e  Y_{e}^{\dagger}  Y_e  Y_{e}^{\dagger}}\Big) +\lambda_5 \mbox{Tr}\Big({Y_e  Y_{e}^{\dagger}  Y_{N}^{T}  Y_{N}^*}\Big) -\frac{1}{2} \lambda_5 \mbox{Tr}\Big({Y_{N}  Y_{N}^{\dagger}  Y_{N}  Y_{N}^{\dagger}}\Big) -\frac{3}{2} \lambda_5 \mbox{Tr}\Big({Y_u  Y_{u}^{\dagger}  Y_u  Y_{u}^{\dagger}}\Big) \Bigg] \, .  \\
	\end{align*}
	\subsection{Gauge Couplings}
	\footnotesize{
		\begingroup
		\allowdisplaybreaks
		\begin{align*}
			\beta_{g_1} \  = \ &  
			\frac{1}{16\pi^2}\Bigg[\frac{21}{5} g_{1}^{3}\Bigg]+\frac{1}{(16\pi^2)^2}\Bigg[ \frac{1}{50} g_{1}^{3} \Big(-15 \mbox{Tr}\Big({Y_{N}  Y_{N}^{\dagger}}\Big)  + 180 g_{2}^{2}  + 208 g_{1}^{2}  -25 \mbox{Tr}\Big({Y_d  Y_{d}^{\dagger}}\Big)  + 440 g_{3}^{2}  -75 \mbox{Tr}\Big({Y_e  Y_{e}^{\dagger}}\Big)  -85 \mbox{Tr}\Big({Y_u  Y_{u}^{\dagger}}\Big) \Big)\Bigg] \, .  \\
			\beta_{g_2} \  = \ &  
			\frac{1}{16\pi^2}\Bigg[-3 g_{2}^{3}\Bigg]+\frac{1}{(16\pi^2)^2}\Bigg[ \frac{1}{10} g_{2}^{3} \Big(120 g_{3}^{2}  + 12 g_{1}^{2}  -15 \mbox{Tr}\Big({Y_d  Y_{d}^{\dagger}}\Big)  -15 \mbox{Tr}\Big({Y_u  Y_{u}^{\dagger}}\Big)  -5 \mbox{Tr}\Big({Y_e  Y_{e}^{\dagger}}\Big)  -5 \mbox{Tr}\Big({Y_{N}  Y_{N}^{\dagger}}\Big)  + 80 g_{2}^{2} \Big)\Bigg] \, .  \\
			\beta_{g_3} \  = \ &  
			\frac{1}{16\pi^2}\Bigg[-7 g_{3}^{3}\Bigg]+\frac{1}{(16\pi^2)^2}\Bigg[-\frac{1}{10} g_{3}^{3} \Big(-11 g_{1}^{2}  + 20 \mbox{Tr}\Big({Y_d  Y_{d}^{\dagger}}\Big)  + 20 \mbox{Tr}\Big({Y_u  Y_{u}^{\dagger}}\Big)  + 260 g_{3}^{2}  -45 g_{2}^{2} \Big)\Bigg] \, .  \\
		\end{align*}
		\endgroup

\section{Dimensional reduction} \label{dim}
The scalar potential in 3-D effective theory is written as follows;
\begin{eqnarray}
	\rm  V_{scalar} &&= m_{11,3}^2\Phi_1^\dagger \Phi_1 + m_{22,3}^2\Phi_2^\dagger\Phi_2 + \lambda_{1,3}(\Phi_1^\dagger \Phi_1)^2 + \lambda_{2,3}(\Phi_2^\dagger \Phi_2)^2 +
	\lambda_{3,3}(\Phi_1^\dagger \Phi_1)(\Phi_2^\dagger \Phi_2) \nn \\ &&+ \lambda_{4,3}(\Phi_1^\dagger \Phi_2)(\Phi_2^\dagger \Phi_1) + [\frac{\lambda_{5,3}}{2}((\Phi_1^\dagger \Phi_2)^2) + h.c].  \label{Eq:2.3}
\end{eqnarray}
where "3" in the subscript for each parameter in the scalar potential denotes the 3-D EFT couplings. IN order to cancel out the scale dependence at one-loop level in one-loop effective potential, all the parameters in the theory will become scale dependent in reduction from 4-D to 3-D and the expressions are as follows; 
\bea
\lambda_{1,3} & = & T\Big[\lambda_1 (\Lambda) + \frac{1}{(4\pi)^2}\Big(\frac{2-3L_b}{16}(3g_2^4+2g_2^2g_1^2+g_1^4)+ N_c L_f(y_t^4-2\lambda_1y_t^2)+L_b\Big(\frac{3}{2}(3g_2^2+g_1^2)\lambda_1-12\lambda_1^2-\lambda_3^2-\lambda_3\lambda_4-\frac{1}{2}\lambda_4^2-\frac{1}{2}\lambda_5^2\Big)\Big)\Big] \nn ,\\
\lambda_{2,3} & = & T\Big[\lambda_2(\Lambda)+ \frac{1}{(4\pi)^2}\Big(\frac{2-3L_b}{16}(3g_2^4+2g_2^2g_1^2+g_1^4)+L_b\Big(\frac{3}{2}(3g_2^2+g_1^2)\lambda_1-12\lambda_2^2-\lambda_3^2-\lambda_3\lambda_4-\frac{1}{2}\lambda_4^2-\frac{1}{2}\lambda_5^2\Big)\Big)\Big], \nn\\
\lambda_{3,3} & = & T\Big[\lambda_{3}(\Lambda)+ \frac{1}{(4\pi)^2}\Big(\frac{2-3L_b}{8}(3g_2^4-2g_2^2g_1^2+g_1^4)-3L_f\lambda_3y_t^2+L_b\Big(\frac{3}{2}(3g_2^2+g_1^2)\lambda_3-6\lambda_1\lambda_3-2\lambda_1\lambda_4-6\lambda_2\lambda_3-2\lambda_2\lambda_4-2\lambda_3^2  \nn\\
&&-\lambda_4^2 -\lambda_5^2\Big)\Big)\Big], \nn\\
\lambda_{4,3} & = & T\Big[\lambda_{4}(\Lambda)+\frac{1}{(4\pi)^2}(g_1^2g_2^2 -3L_f \lambda_4y_t^2 -L_b(\frac{3}{2}g_2^2g_1^2+2\lambda_1\lambda_4+2\lambda_2\lambda_4+2\lambda_4^2+4\lambda_3\lambda_4+4\lambda_5^2-\frac{3}{2}(3g_2^2+g_1^2)\lambda_4) )\Big],\nn\\
\lambda_{5,3} & = & T\Big[\lambda_{5}(\Lambda)+\frac{1}{(4\pi)^2}(-3L_f\lambda_5 y_t^2-L_b(2\lambda_1\lambda_5+2\lambda_2\lambda_5+4\lambda_3\lambda_5 + 6\lambda_4\lambda_5-\frac{3}{2}(3g_2^2+g_1^2)\lambda_5 ))\Big],
\eea
where 
\bea
L_b & = & \ln\Big(\frac{\Lambda^2}{T^2}\Big)-2[\ln(4\pi)-\gamma],\\
L_f & = & L_b + 4 \ln2.
\eea
Here, $L_b$ and $L_f$ are logarithms that arise frequently from one-loop bosonic and fermionic sum integrals with $\Lambda$ is the $\overline{MS}$ scale and $\gamma$ is the Euler-Mascheroni constant. The expressions for the mass parameters are computed as follows:
where
\bea
(m_{11,3}^2)_{\rm IDM} & = & (m_{11,3}^2)_{\rm SM} + \frac{T^2}{12}\Big(2\lambda_3(\Lambda)+\lambda_4(\Lambda)\Big) + \frac{1}{16\pi^2}\Big[m_{22}^2\Big(-L_b(2\lambda_3+\lambda_4)\Big)+T^2\Big(\frac{5}{48}g_2^4+\frac{5}{144}g_1^4 +\frac{1}{24}(3g_2^2+g_1^2) \nn \\
&& (2\lambda_3+\lambda_4)+\frac{1}{T^2}\Big(c+ln(\frac{3T}{\Lambda_3})\Big)\Big(-\frac{1}{8}(3g_{2,3}^4+g_{1,3}^4)+\frac{1}{2}(3g_{2,3}^2+g_{1,3}^2)(2\lambda_{3,3}+\lambda_{4,3})-2(\lambda_{3,3}^2+\lambda_{3,3}\lambda_{4,3}+\lambda_{4,3}^2)\nn \\
&&-3\lambda_{5,3}^2\Big) + L_b\Big(-\frac{7}{32}g_2^4-\frac{7}{96}g_1^4-\frac{1}{2}(\lambda_1+\lambda_2)(2\lambda_3+\lambda_4) -\frac{5}{6}\lambda_3^2 -\frac{7}{12}\lambda_4^2 -\frac{5}{6}\lambda-3\lambda_4 -\frac{3}{4}\lambda_5^2 +\frac{1}{8}(3 g_2^2+g_1^2)(2\lambda_3 +\lambda_4)\Big)\nn \\
&&+\Big(-\frac{1}{4}y_t^2(2\lambda_3+\lambda_4)\Big)L_f\Big) \Big],\\
(m_{22,3}^2)_{\rm IDM} & = & m_{22}^2(\Lambda) + \frac{T^2}{16}\Big(3g_2^2(\Lambda)+g_1^2(\Lambda)+8\lambda_2(\Lambda)+\frac{4}{3}(2\lambda_3(\Lambda)+\lambda_4(\Lambda))\Big)+ \frac{1}{16\pi^2}\Big[L_b\Big(\frac{3}{4}(3g_2^2+g_1^2)-6\lambda_2)m_{22}^2\nn\\
&& -(2\lambda_3+\lambda_4)m_{11}^2\Big)+T^2\Big(\frac{59}{32}g_2^4 +\frac{11}{288}g_1^4-\frac{3}{16}g_2^2g_1^2+\frac{1}{4}\lambda_1(3g_2^2+g_1^2)+\frac{1}{24}(3g_2^2+g_1^2)(2\lambda_3+\lambda_4)+L_b\Big(\frac{27}{32}g_2^4 \nn\\
&&-\frac{17}{96}g_1^4-\frac{3}{16}g_2^2g_1^2+\frac{1}{8}(3g_2^2+g_1^2)(6\lambda_1+2\lambda_3+\lambda_4)-\frac{1}{2}(\lambda_1+\lambda_2)(2\lambda_3+\lambda_4)-6\lambda_1^2-\frac{5}{6}\lambda_3^2-\frac{5}{6}\lambda_3\lambda_4 \nn\\
&&-\frac{7}{12}\lambda_4^2-\frac{3}{4}\lambda_5^2\Big) +\frac{1}{T^2}\Big(c+\ln(\frac{3T}{\Lambda_3})\Big)\Big(\frac{33}{16}g_{2,3}^4+12g_{2,3}^2h_1-6h_1^2+9g_{2,3}^2\lambda_{1,3}-12\lambda_{1,3}^2 -\frac{7}{16}g_{1,3}^4-\frac{9}{8}g_{2,3}^2g_{1,3}^2\nn \\
&&-2h_2^2-3h_3^2+3g_{1,3}^2\lambda_{1,3}+\frac{1}{2}(3g_{2,3}^2+g_{1,3}^2)(2\lambda_{3,3}+\lambda_{4,3})-2(\lambda_{3,3}^2+\lambda_{3,3}\lambda_{4,3}+\lambda_{4,3}^2)-3\lambda_{5,3}^2\Big)+(\frac{1}{12}g_2^4\nn \\
&&+\frac{5}{108}g_1^4)n_f+L_f(-\frac{1}{4}y_t^2(2\lambda_3+\lambda_4)-(\frac{1}{4}g_2^4+\frac{5}{36}g_1^4)n_f)+\ln(2)(\frac{3}{2}y_t^2(2\lambda_3+\lambda_4)+(\frac{3}{2}g_2^4+\frac{5}{6}g_1^4)n_f)\Big)\Big], \\
(m_{11,3}^2)_{\rm SM} & = & - m_{11}^2(\Lambda) + \frac{T^2}{12}\Big(\frac{3}{4}(3g_2^2(\Lambda)+g_1^2(\Lambda))+N_cy_t^2(\Lambda)+8\lambda_1(\Lambda)\Big)+ \frac{m_{11}^2(\Lambda)}{(4\pi)^2}\Big(\Big(\frac{3}{4}(3g_2^2+g_1^2)-6\lambda_1\Big)L_b-N_cy_t^2L_f\Big) \nn \\
&& + \frac{T^2}{(4\pi)^2}\Big[\frac{167}{96}g_2^4 + \frac{1}{288}g_1^4 -\frac{3}{16}g_2^2g_1^2 + \frac{(1+3L_b)}{4}\lambda_1(3g_2^2+g_1^2)+L_b\Big(\frac{17}{16}g_2^4 - \frac{5}{48}g_1^4-\frac{3}{16}g_2^2g_1^2-6\lambda_1^2\Big) \nn \\
&& +\frac{1}{T^2}\Big(c+ \ln \Big(\frac{3T}{\Lambda_{3d}}\Big)\Big)\Big(\frac{39}{16}g_{2,3}^4+ 12 g_{2,3}^2h_4 - 6 h_4^2 + 9 g_{2,3}^2\lambda_{1,3}-12\lambda_{1,3}^2-\frac{5}{16}g_{1,3}^4 -\frac{9}{8}g_{2,3}^2g_{1,3}^2-2h_5^2-3h_6^2 \nn \\
&& +3g_{1,3}^2\lambda_{1,3}\Big) -\frac{1}{96}\Big(9L_b-3L_f-2\Big)\Big((N_c+1)g_2^4 + \frac{1}{6}Y_{2f}g_1^4\Big)n_f + \frac{N_c}{32}\Big(7L_b-L_f-2\Big)g_2^2y_t^2 \nn \\
&& -\frac{N_c}{4}(3L_b+L_f)\lambda_1y_t^2 + \frac{N_c}{96}\Big(\Big(9(L_b-L_f)+4\Big)Y_{\phi}^2-2\Big(L_b-4L_f+3\Big)(Y_q^2+Y_u^2)\Big)g_1^2 y_t^2 \nn \\
&& -\frac{N_cC_F}{6}\Big(L_b-4L_f+3\Big)g_s^2y_t^2+\frac{N_c}{24}\Big(3L_b-2(N_c-3)L_f\Big)y_t^4\Big].
\eea
with $C_F=\frac{N_c^2-1}{2N_c}=\frac{4}{3}, Y_{\phi}=1, Y_{2f}=\frac{40}{3}$ and $c \sim -0.348723.$ is the fundamental quadratic Casimir of $SU(3)$. The gauge couplings and the other relevant couplings are given as follows;
\bea
g_{1,3}^2 & = & g_1^2(\Lambda)T\Big(1+\frac{g_1^2}{(4\pi)^2}\Big[-\frac{N_d}{6}L_b-\frac{20}{9}n_fL_f\Big]\Big), \nn \\
g_{2,3}^2 & = & g_2^2(\Lambda)T\Big(1+\frac{g_2^2}{(4\pi)^2}\Big[\frac{44-N_d}{6}L_b+\frac{2}{3}-\frac{4}{3}n_fL_f\Big]\Big), \nn \\
h_4 & = & \frac{g_2^2(\Lambda)T}{4}\Big(1+\frac{1}{(4\pi)^2}\Big[\Big(\frac{44-N_d}{6}L_b+\frac{53}{6}-\frac{N_d}{3}-\frac{4n_f}{3}(L_f-1)\Big)g_2^2+\frac{g_1^2}{2}-6y_t^2+12\lambda_2+2(2\lambda_3+\lambda_4)\Big]\Big), \nn \\
h_5 & = & \frac{g_1^2(\Lambda)T}{4}\Big(1+\frac{1}{(4\pi)^2}\Big[\frac{3}{2}g_2^2 +\Big(\frac{1}{2}-\frac{N_d}{6}(2+L_b)-\frac{20}{9}n_f(L_f-1)\Big)g_1^2-\frac{34}{3}y_t^2+12\lambda_2+2(2\lambda_3+\lambda_4)\Big]\Big), \nn \\
h_6 & = & \frac{g_2(\Lambda)g_1(\Lambda)T}{2}\Big(1+\frac{1}{(4\pi)^2}\Big[-\frac{5+N_d}{6}g_2^2+\frac{3-N_d}{6}g_1^2+L_b(\frac{44-N_d}{12}g_2^2-\frac{N_d}{12}g_1^2)-n_f(L_f-1)(\frac{2}{3}g_2^2+\frac{10}{9}g_1^2)+2y_t^2+4\lambda_2 \nn \\
&&+2\lambda_4\Big]\Big),
\eea
with $N_d=1, n_f=3, N_c=3$.

\bibliography{References}

\providecommand{\href}[2]{#2}\begingroup\begin{thebibliography}{10}

\bibitem{Biswas:2025woq}
A.~Biswas, S.~Jangid and S.~C. Park, \textit{{Investigating two-zero texture in
  the light of gauged Type-II seesaw}},
  \href{https://arxiv.org/abs/2504.10312}{{\ttfamily 2504.10312}}.

\bibitem{Jangid:2025ded}
S.~Jangid, A.~Biswas and S.~C. Park, \textit{{Strongly electroweak phase
  transition with $U(1)_{L_{\mu}-L_{\tau}}$ gauged non-zero hypercharge
  triplet}},  \href{https://arxiv.org/abs/2504.10874}{{\ttfamily 2504.10874}}.

\bibitem{Guth:1980zm}
A.~H. Guth, \textit{{The Inflationary Universe: A Possible Solution to the
  Horizon and Flatness Problems}},
  \href{https://doi.org/10.1103/PhysRevD.23.347}{\textit{Phys. Rev. D}
  {\bfseries 23} (1981) 347--356}.

\bibitem{Sher:1988mj}
M.~Sher, \textit{{Electroweak Higgs Potentials and Vacuum Stability}},
  \href{https://doi.org/10.1016/0370-1573(89)90061-6}{\textit{Phys. Rept.}
  {\bfseries 179} (1989) 273--418}.

\bibitem{Buttazzo:2013uya}
D.~Buttazzo, G.~Degrassi, P.~P. Giardino, G.~F. Giudice, F.~Sala, A.~Salvio
  et~al., \textit{{Investigating the near-criticality of the Higgs boson}},
  \href{https://doi.org/10.1007/JHEP12(2013)089}{\textit{JHEP} {\bfseries 12}
  (2013) 089}, [\href{https://arxiv.org/abs/1307.3536}{{\ttfamily 1307.3536}}].

\bibitem{Degrassi:2012ry}
G.~Degrassi, S.~Di~Vita, J.~Elias-Miro, J.~R. Espinosa, G.~F. Giudice,
  G.~Isidori et~al., \textit{{Higgs mass and vacuum stability in the Standard
  Model at NNLO}}, \href{https://doi.org/10.1007/JHEP08(2012)098}{\textit{JHEP}
  {\bfseries 08} (2012) 098},
  [\href{https://arxiv.org/abs/1205.6497}{{\ttfamily 1205.6497}}].

\bibitem{Planck:2015sxf}
{\scshape Planck} collaboration, P.~A.~R. Ade et~al., \textit{{Planck 2015
  results. XX. Constraints on inflation}},
  \href{https://doi.org/10.1051/0004-6361/201525898}{\textit{Astron.
  Astrophys.} {\bfseries 594} (2016) A20},
  [\href{https://arxiv.org/abs/1502.02114}{{\ttfamily 1502.02114}}].

\bibitem{WMAP:2010qai}
{\scshape WMAP} collaboration, E.~Komatsu et~al., \textit{{Seven-Year Wilkinson
  Microwave Anisotropy Probe (WMAP) Observations: Cosmological
  Interpretation}},
  \href{https://doi.org/10.1088/0067-0049/192/2/18}{\textit{Astrophys. J.
  Suppl.} {\bfseries 192} (2011) 18},
  [\href{https://arxiv.org/abs/1001.4538}{{\ttfamily 1001.4538}}].

\bibitem{Lerner:2009na}
R.~N. Lerner and J.~McDonald, \textit{{Higgs Inflation and Naturalness}},
  \href{https://doi.org/10.1088/1475-7516/2010/04/015}{\textit{JCAP} {\bfseries
  04} (2010) 015}, [\href{https://arxiv.org/abs/0912.5463}{{\ttfamily
  0912.5463}}].

\bibitem{Lerner:2011ge}
R.~N. Lerner and J.~McDonald, \textit{{Distinguishing Higgs inflation and its
  variants}}, \href{https://doi.org/10.1103/PhysRevD.83.123522}{\textit{Phys.
  Rev. D} {\bfseries 83} (2011) 123522},
  [\href{https://arxiv.org/abs/1104.2468}{{\ttfamily 1104.2468}}].

\bibitem{Linde:1983gd}
A.~D. Linde, \textit{{Chaotic Inflation}},
  \href{https://doi.org/10.1016/0370-2693(83)90837-7}{\textit{Phys. Lett. B}
  {\bfseries 129} (1983) 177--181}.

\bibitem{Linde:2014nna}
A.~Linde, \textit{{Inflationary Cosmology after Planck 2013}},  in
  \textit{{100e Ecole d'Ete de Physique: Post-Planck Cosmology}}, pp.~231--316,
  2015, \href{https://arxiv.org/abs/1402.0526}{{\ttfamily 1402.0526}},
  \href{https://doi.org/10.1093/acprof:oso/9780198728856.003.0006}{DOI}.

\bibitem{Trodden:1998ym}
M.~Trodden, \textit{{Electroweak baryogenesis}},
  \href{https://doi.org/10.1103/RevModPhys.71.1463}{\textit{Rev. Mod. Phys.}
  {\bfseries 71} (1999) 1463--1500},
  [\href{https://arxiv.org/abs/hep-ph/9803479}{{\ttfamily hep-ph/9803479}}].

\bibitem{Cohen:1993nk}
A.~G. Cohen, D.~B. Kaplan and A.~E. Nelson, \textit{{Progress in electroweak
  baryogenesis}},
  \href{https://doi.org/10.1146/annurev.ns.43.120193.000331}{\textit{Ann. Rev.
  Nucl. Part. Sci.} {\bfseries 43} (1993) 27--70},
  [\href{https://arxiv.org/abs/hep-ph/9302210}{{\ttfamily hep-ph/9302210}}].

\bibitem{Morrissey:2012db}
D.~E. Morrissey and M.~J. Ramsey-Musolf, \textit{{Electroweak baryogenesis}},
  \href{https://doi.org/10.1088/1367-2630/14/12/125003}{\textit{New J. Phys.}
  {\bfseries 14} (2012) 125003},
  [\href{https://arxiv.org/abs/1206.2942}{{\ttfamily 1206.2942}}].

\bibitem{Baker:2021zsf}
M.~J. Baker, M.~Breitbach, J.~Kopp, L.~Mittnacht and Y.~Soreq,
  \textit{{Filtered baryogenesis}},
  \href{https://doi.org/10.1007/JHEP08(2022)010}{\textit{JHEP} {\bfseries 08}
  (2022) 010}, [\href{https://arxiv.org/abs/2112.08987}{{\ttfamily
  2112.08987}}].

\bibitem{Cline:2020jre}
J.~M. Cline and K.~Kainulainen, \textit{{Electroweak baryogenesis at high
  bubble wall velocities}},
  \href{https://doi.org/10.1103/PhysRevD.101.063525}{\textit{Phys. Rev. D}
  {\bfseries 101} (2020) 063525},
  [\href{https://arxiv.org/abs/2001.00568}{{\ttfamily 2001.00568}}].

\bibitem{Jangid:2023jya}
S.~Jangid and H.~Okada, \textit{{Exploring CP-violation in Y=0 inert triplet
  with real singlet}},
  \href{https://doi.org/10.1103/PhysRevD.108.055025}{\textit{Phys. Rev. D}
  {\bfseries 108} (2023) 055025},
  [\href{https://arxiv.org/abs/2304.13325}{{\ttfamily 2304.13325}}].

\bibitem{PhysRevD.30.2212}
F.~R. Klinkhamer and N.~S. Manton, \textit{A saddle-point solution in the
  weinberg-salam theory},
  \href{https://doi.org/10.1103/PhysRevD.30.2212}{\textit{Phys. Rev. D}
  {\bfseries 30} (Nov, 1984) 2212--2220}.

\bibitem{Gavela:1994dt}
M.~B. Gavela, P.~Hernandez, J.~Orloff, O.~Pene and C.~Quimbay,
  \textit{{Standard model CP violation and baryon asymmetry. Part 2: Finite
  temperature}},
  \href{https://doi.org/10.1016/0550-3213(94)00410-2}{\textit{Nucl. Phys. B}
  {\bfseries 430} (1994) 382--426},
  [\href{https://arxiv.org/abs/hep-ph/9406289}{{\ttfamily hep-ph/9406289}}].

\bibitem{Huet:1994jb}
P.~Huet and E.~Sather, \textit{{Electroweak baryogenesis and standard model CP
  violation}}, \href{https://doi.org/10.1103/PhysRevD.51.379}{\textit{Phys.
  Rev. D} {\bfseries 51} (1995) 379--394},
  [\href{https://arxiv.org/abs/hep-ph/9404302}{{\ttfamily hep-ph/9404302}}].

\bibitem{Aoki:1999fi}
Y.~Aoki, F.~Csikor, Z.~Fodor and A.~Ukawa, \textit{{The Endpoint of the first
  order phase transition of the SU(2) gauge Higgs model on a four-dimensional
  isotropic lattice}},
  \href{https://doi.org/10.1103/PhysRevD.60.013001}{\textit{Phys. Rev. D}
  {\bfseries 60} (1999) 013001},
  [\href{https://arxiv.org/abs/hep-lat/9901021}{{\ttfamily hep-lat/9901021}}].

\bibitem{Kajantie:1996mn}
K.~Kajantie, M.~Laine, K.~Rummukainen and M.~E. Shaposhnikov, \textit{{Is there
  a~ hot electroweak phase transition at $m_H \gtrsim m_W$?}},
  \href{https://doi.org/10.1103/PhysRevLett.77.2887}{\textit{Phys. Rev. Lett.}
  {\bfseries 77} (1996) 2887--2890},
  [\href{https://arxiv.org/abs/hep-ph/9605288}{{\ttfamily hep-ph/9605288}}].

\bibitem{Kajantie:1996qd}
K.~Kajantie, M.~Laine, K.~Rummukainen and M.~E. Shaposhnikov, \textit{{A
  Nonperturbative analysis of the finite T phase transition in SU(2) x U(1)
  electroweak theory}},
  \href{https://doi.org/10.1016/S0550-3213(97)00164-8}{\textit{Nucl. Phys. B}
  {\bfseries 493} (1997) 413--438},
  [\href{https://arxiv.org/abs/hep-lat/9612006}{{\ttfamily hep-lat/9612006}}].

\bibitem{Csikor:1998eu}
F.~Csikor, Z.~Fodor and J.~Heitger, \textit{{Endpoint of the hot electroweak
  phase transition}},
  \href{https://doi.org/10.1103/PhysRevLett.82.21}{\textit{Phys. Rev. Lett.}
  {\bfseries 82} (1999) 21--24},
  [\href{https://arxiv.org/abs/hep-ph/9809291}{{\ttfamily hep-ph/9809291}}].

\bibitem{Kajantie:1995kf}
K.~Kajantie, M.~Laine, K.~Rummukainen and M.~E. Shaposhnikov, \textit{{The
  Electroweak phase transition: A Nonperturbative analysis}},
  \href{https://doi.org/10.1016/0550-3213(96)00052-1}{\textit{Nucl. Phys. B}
  {\bfseries 466} (1996) 189--258},
  [\href{https://arxiv.org/abs/hep-lat/9510020}{{\ttfamily hep-lat/9510020}}].

\bibitem{Hindmarsh:2015qta}
M.~Hindmarsh, S.~J. Huber, K.~Rummukainen and D.~J. Weir, \textit{{Numerical
  simulations of acoustically generated gravitational waves at a first order
  phase transition}},
  \href{https://doi.org/10.1103/PhysRevD.92.123009}{\textit{Phys. Rev. D}
  {\bfseries 92} (2015) 123009},
  [\href{https://arxiv.org/abs/1504.03291}{{\ttfamily 1504.03291}}].

\bibitem{LIGOScientific:2016aoc}
{\scshape LIGO Scientific, Virgo} collaboration, B.~P. Abbott et~al.,
  \textit{{Observation of Gravitational Waves from a Binary Black Hole
  Merger}}, \href{https://doi.org/10.1103/PhysRevLett.116.061102}{\textit{Phys.
  Rev. Lett.} {\bfseries 116} (2016) 061102},
  [\href{https://arxiv.org/abs/1602.03837}{{\ttfamily 1602.03837}}].

\bibitem{LIGOScientific:2016dsl}
{\scshape LIGO Scientific, Virgo} collaboration, B.~P. Abbott et~al.,
  \textit{{Binary Black Hole Mergers in the first Advanced LIGO Observing
  Run}}, \href{https://doi.org/10.1103/PhysRevX.6.041015}{\textit{Phys. Rev. X}
  {\bfseries 6} (2016) 041015},
  [\href{https://arxiv.org/abs/1606.04856}{{\ttfamily 1606.04856}}]. [Erratum:
  Phys.Rev.X 8, 039903 (2018)].

\bibitem{eLISA:2013xep}
{\scshape eLISA} collaboration, P.~A. Seoane et~al., \textit{{The Gravitational
  Universe}},  \href{https://arxiv.org/abs/1305.5720}{{\ttfamily 1305.5720}}.

\bibitem{Lerner:2009xg}
R.~N. Lerner and J.~McDonald, \textit{{Gauge singlet scalar as inflaton and
  thermal relic dark matter}},
  \href{https://doi.org/10.1103/PhysRevD.80.123507}{\textit{Phys. Rev. D}
  {\bfseries 80} (2009) 123507},
  [\href{https://arxiv.org/abs/0909.0520}{{\ttfamily 0909.0520}}].

\bibitem{Cline:2012hg}
J.~M. Cline and K.~Kainulainen, \textit{{Electroweak baryogenesis and dark
  matter from a singlet Higgs}},
  \href{https://doi.org/10.1088/1475-7516/2013/01/012}{\textit{JCAP} {\bfseries
  01} (2013) 012}, [\href{https://arxiv.org/abs/1210.4196}{{\ttfamily
  1210.4196}}].

\bibitem{Alanne:2014bra}
T.~Alanne, K.~Tuominen and V.~Vaskonen, \textit{{Strong phase transition, dark
  matter and vacuum stability from simple hidden sectors}},
  \href{https://doi.org/10.1016/j.nuclphysb.2014.11.001}{\textit{Nucl. Phys. B}
  {\bfseries 889} (2014) 692--711},
  [\href{https://arxiv.org/abs/1407.0688}{{\ttfamily 1407.0688}}].

\bibitem{Choubey:2017hsq}
S.~Choubey and A.~Kumar, \textit{{Inflation and Dark Matter in the Inert
  Doublet Model}}, \href{https://doi.org/10.1007/JHEP11(2017)080}{\textit{JHEP}
  {\bfseries 11} (2017) 080},
  [\href{https://arxiv.org/abs/1707.06587}{{\ttfamily 1707.06587}}].

\bibitem{Benincasa:2022elt}
N.~Benincasa, L.~Delle~Rose, K.~Kannike and L.~Marzola, \textit{{Multi-step
  phase transitions and gravitational waves in the inert doublet model}},
  \href{https://doi.org/10.1088/1475-7516/2022/12/025}{\textit{JCAP} {\bfseries
  12} (2022) 025}, [\href{https://arxiv.org/abs/2205.06669}{{\ttfamily
  2205.06669}}].

\bibitem{Fabian:2020hny}
S.~Fabian, F.~Goertz and Y.~Jiang, \textit{{Dark matter and nature of
  electroweak phase transition with an inert doublet}},
  \href{https://doi.org/10.1088/1475-7516/2021/09/011}{\textit{JCAP} {\bfseries
  09} (2021) 011}, [\href{https://arxiv.org/abs/2012.12847}{{\ttfamily
  2012.12847}}].

\bibitem{Majumdar:2020vdd}
D.~Majumdar, A.~Paul and B.~Banerjee, \textit{{Signatures of GW from an
  Extended Inert Doublet Model}},
  \href{https://doi.org/10.1007/978-981-15-6292-1_22}{\textit{Springer Proc.
  Phys.} {\bfseries 248} (2020) 183--191}.

\bibitem{Blinov:2015vma}
N.~Blinov, S.~Profumo and T.~Stefaniak, \textit{{The Electroweak Phase
  Transition in the Inert Doublet Model}},
  \href{https://doi.org/10.1088/1475-7516/2015/07/028}{\textit{JCAP} {\bfseries
  07} (2015) 028}, [\href{https://arxiv.org/abs/1504.05949}{{\ttfamily
  1504.05949}}].

\bibitem{Borah:2012pu}
D.~Borah and J.~M. Cline, \textit{{Inert Doublet Dark Matter with Strong
  Electroweak Phase Transition}},
  \href{https://doi.org/10.1103/PhysRevD.86.055001}{\textit{Phys. Rev. D}
  {\bfseries 86} (2012) 055001},
  [\href{https://arxiv.org/abs/1204.4722}{{\ttfamily 1204.4722}}].

\bibitem{Jangid:2020dqh}
S.~Jangid, P.~Bandyopadhyay, P.~S. Bhupal~Dev and A.~Kumar, \textit{{Vacuum
  stability in inert higgs doublet model with right-handed neutrinos}},
  \href{https://doi.org/10.1007/JHEP08(2020)154}{\textit{JHEP} {\bfseries 08}
  (2020) 154}, [\href{https://arxiv.org/abs/2001.01764}{{\ttfamily
  2001.01764}}].

\bibitem{Espinosa:2007qp}
J.~R. Espinosa, G.~F. Giudice and A.~Riotto, \textit{{Cosmological implications
  of the Higgs mass measurement}},
  \href{https://doi.org/10.1088/1475-7516/2008/05/002}{\textit{JCAP} {\bfseries
  05} (2008) 002}, [\href{https://arxiv.org/abs/0710.2484}{{\ttfamily
  0710.2484}}].

\bibitem{Espinosa:2015qea}
J.~R. Espinosa, G.~F. Giudice, E.~Morgante, A.~Riotto, L.~Senatore, A.~Strumia
  et~al., \textit{{The cosmological Higgstory of the vacuum instability}},
  \href{https://doi.org/10.1007/JHEP09(2015)174}{\textit{JHEP} {\bfseries 09}
  (2015) 174}, [\href{https://arxiv.org/abs/1505.04825}{{\ttfamily
  1505.04825}}].

\bibitem{Kobakhidze:2013tn}
A.~Kobakhidze and A.~Spencer-Smith, \textit{{Electroweak Vacuum (In)Stability
  in an Inflationary Universe}},
  \href{https://doi.org/10.1016/j.physletb.2013.04.013}{\textit{Phys. Lett. B}
  {\bfseries 722} (2013) 130--134},
  [\href{https://arxiv.org/abs/1301.2846}{{\ttfamily 1301.2846}}].

\bibitem{Enqvist:2013kaa}
K.~Enqvist, T.~Meriniemi and S.~Nurmi, \textit{{Generation of the Higgs
  Condensate and Its Decay after Inflation}},
  \href{https://doi.org/10.1088/1475-7516/2013/10/057}{\textit{JCAP} {\bfseries
  10} (2013) 057}, [\href{https://arxiv.org/abs/1306.4511}{{\ttfamily
  1306.4511}}].

\bibitem{Fairbairn:2014zia}
M.~Fairbairn and R.~Hogan, \textit{{Electroweak Vacuum Stability in light of
  BICEP2}}, \href{https://doi.org/10.1103/PhysRevLett.112.201801}{\textit{Phys.
  Rev. Lett.} {\bfseries 112} (2014) 201801},
  [\href{https://arxiv.org/abs/1403.6786}{{\ttfamily 1403.6786}}].

\bibitem{Enqvist:2014bua}
K.~Enqvist, T.~Meriniemi and S.~Nurmi, \textit{{Higgs Dynamics during
  Inflation}},
  \href{https://doi.org/10.1088/1475-7516/2014/07/025}{\textit{JCAP} {\bfseries
  07} (2014) 025}, [\href{https://arxiv.org/abs/1404.3699}{{\ttfamily
  1404.3699}}].

\bibitem{Kobakhidze:2014xda}
A.~Kobakhidze and A.~Spencer-Smith, \textit{{The Higgs vacuum is unstable}},
  \href{https://arxiv.org/abs/1404.4709}{{\ttfamily 1404.4709}}.

\bibitem{Herranen:2014cua}
M.~Herranen, T.~Markkanen, S.~Nurmi and A.~Rajantie, \textit{{Spacetime
  curvature and the Higgs stability during inflation}},
  \href{https://doi.org/10.1103/PhysRevLett.113.211102}{\textit{Phys. Rev.
  Lett.} {\bfseries 113} (2014) 211102},
  [\href{https://arxiv.org/abs/1407.3141}{{\ttfamily 1407.3141}}].

\bibitem{Kamada:2014ufa}
K.~Kamada, \textit{{Inflationary cosmology and the standard model Higgs with a
  small Hubble induced mass}},
  \href{https://doi.org/10.1016/j.physletb.2015.01.024}{\textit{Phys. Lett. B}
  {\bfseries 742} (2015) 126--135},
  [\href{https://arxiv.org/abs/1409.5078}{{\ttfamily 1409.5078}}].

\bibitem{Shkerin:2015exa}
A.~Shkerin and S.~Sibiryakov, \textit{{On stability of electroweak vacuum
  during inflation}},
  \href{https://doi.org/10.1016/j.physletb.2015.05.012}{\textit{Phys. Lett. B}
  {\bfseries 746} (2015) 257--260},
  [\href{https://arxiv.org/abs/1503.02586}{{\ttfamily 1503.02586}}].

\bibitem{Isidori:2001bm}
G.~Isidori, G.~Ridolfi and A.~Strumia, \textit{{On the metastability of the
  standard model vacuum}},
  \href{https://doi.org/10.1016/S0550-3213(01)00302-9}{\textit{Nucl. Phys. B}
  {\bfseries 609} (2001) 387--409},
  [\href{https://arxiv.org/abs/hep-ph/0104016}{{\ttfamily hep-ph/0104016}}].

\bibitem{DelleRose:2015bpo}
L.~Delle~Rose, C.~Marzo and A.~Urbano, \textit{{On the fate of the Standard
  Model at finite temperature}},
  \href{https://doi.org/10.1007/JHEP05(2016)050}{\textit{JHEP} {\bfseries 05}
  (2016) 050}, [\href{https://arxiv.org/abs/1507.06912}{{\ttfamily
  1507.06912}}].

\bibitem{PhysRevD.44.3620}
P.~Arnold and S.~Vokos, \textit{Instability of hot electroweak theory: Bounds
  on ${\mathit{m}}_{\mathit{h}}$ and ${\mathit{m}}_{\mathit{t}}$},
  \href{https://doi.org/10.1103/PhysRevD.44.3620}{\textit{Phys. Rev. D}
  {\bfseries 44} (Dec, 1991) 3620--3627}.

\bibitem{Jangid:2023lny}
S.~Jangid and H.~Okada, \textit{{Electroweak phase transition with radiative
  symmetry breaking in a type-II seesaw model with an inert doublet}},
  \href{https://doi.org/10.1103/PhysRevD.109.015001}{\textit{Phys. Rev. D}
  {\bfseries 109} (2024) 015001},
  [\href{https://arxiv.org/abs/2310.12591}{{\ttfamily 2310.12591}}].

\bibitem{Coleman:1977py}
S.~R. Coleman, \textit{{The Fate of the False Vacuum. 1. Semiclassical
  Theory}}, \href{https://doi.org/10.1103/PhysRevD.16.1248}{\textit{Phys. Rev.
  D} {\bfseries 15} (1977) 2929--2936}. [Erratum: Phys.Rev.D 16, 1248 (1977)].

\bibitem{Callan:1977pt}
C.~G. Callan, Jr. and S.~R. Coleman, \textit{{The Fate of the False Vacuum. 2.
  First Quantum Corrections}},
  \href{https://doi.org/10.1103/PhysRevD.16.1762}{\textit{Phys. Rev. D}
  {\bfseries 16} (1977) 1762--1768}.

\bibitem{Anderson:1990aa}
G.~W. Anderson, \textit{{New Cosmological Constraints on the Higgs Boson and
  Top Quark Masses}},
  \href{https://doi.org/10.1016/0370-2693(90)90849-2}{\textit{Phys. Lett. B}
  {\bfseries 243} (1990) 265--270}.

\bibitem{Espinosa:1995se}
J.~R. Espinosa and M.~Quiros, \textit{{Improved metastability bounds on the
  standard model Higgs mass}},
  \href{https://doi.org/10.1016/0370-2693(95)00572-3}{\textit{Phys. Lett. B}
  {\bfseries 353} (1995) 257--266},
  [\href{https://arxiv.org/abs/hep-ph/9504241}{{\ttfamily hep-ph/9504241}}].

\bibitem{LopezHonorez:2010eeh}
L.~Lopez~Honorez and C.~E. Yaguna, \textit{{The inert doublet model of dark
  matter revisited}},
  \href{https://doi.org/10.1007/JHEP09(2010)046}{\textit{JHEP} {\bfseries 09}
  (2010) 046}, [\href{https://arxiv.org/abs/1003.3125}{{\ttfamily 1003.3125}}].

\bibitem{Gustafsson:2010zz}
M.~Gustafsson, \textit{{The Inert Doublet Model and Its Phenomenology}},
  \href{https://doi.org/10.22323/1.114.0030}{\textit{PoS} {\bfseries
  CHARGED2010} (2010) 030}, [\href{https://arxiv.org/abs/1106.1719}{{\ttfamily
  1106.1719}}].

\bibitem{Treesukrat:2019ahh}
W.~Treesukrat and P.~Uttayarat, \textit{{Dark matter from the inert Higgs
  doublet model}},
  \href{https://doi.org/10.1088/1742-6596/1380/1/012093}{\textit{J. Phys. Conf.
  Ser.} {\bfseries 1380} (2019) 012093}.

\bibitem{Goudelis:2013uca}
A.~Goudelis, B.~Herrmann and O.~St\r{a}l, \textit{{Dark matter in the Inert
  Doublet Model after the discovery of a Higgs-like boson at the LHC}},
  \href{https://doi.org/10.1007/JHEP09(2013)106}{\textit{JHEP} {\bfseries 09}
  (2013) 106}, [\href{https://arxiv.org/abs/1303.3010}{{\ttfamily 1303.3010}}].

\bibitem{LopezHonorez:2012zz}
L.~Lopez~Honorez, \textit{{Scalar dark matter: A revision of the inert doublet
  model}}, \href{https://doi.org/10.1393/ncc/i2012-11135-7}{\textit{Nuovo Cim.
  C} {\bfseries 035N1} (2012) 39--46}.

\bibitem{Tytgat:2007cv}
M.~H.~G. Tytgat, \textit{{The Inert Doublet Model: A New archetype of WIMP dark
  matter?}}, \href{https://doi.org/10.1088/1742-6596/120/4/042026}{\textit{J.
  Phys. Conf. Ser.} {\bfseries 120} (2008) 042026},
  [\href{https://arxiv.org/abs/0712.4206}{{\ttfamily 0712.4206}}].

\bibitem{LopezHonorez:2007wm}
L.~Lopez~Honorez, \textit{{Dark Matter from the Inert Doublet Model}},  in
  \textit{{42nd Rencontres de Moriond on Electroweak Interactions and Unified
  Theories}}, pp.~277--282, 6, 2007,
  \href{https://arxiv.org/abs/0706.0186}{{\ttfamily 0706.0186}}.

\bibitem{LopezHonorez:2006gr}
L.~Lopez~Honorez, E.~Nezri, J.~F. Oliver and M.~H.~G. Tytgat, \textit{{The
  Inert Doublet Model: An Archetype for Dark Matter}},
  \href{https://doi.org/10.1088/1475-7516/2007/02/028}{\textit{JCAP} {\bfseries
  02} (2007) 028}, [\href{https://arxiv.org/abs/hep-ph/0612275}{{\ttfamily
  hep-ph/0612275}}].

\bibitem{Bezrukov:2007ep}
F.~L. Bezrukov and M.~Shaposhnikov, \textit{{The Standard Model Higgs boson as
  the inflaton}},
  \href{https://doi.org/10.1016/j.physletb.2007.11.072}{\textit{Phys. Lett. B}
  {\bfseries 659} (2008) 703--706},
  [\href{https://arxiv.org/abs/0710.3755}{{\ttfamily 0710.3755}}].

\bibitem{Park:2008hz}
S.~C. Park and S.~Yamaguchi, \textit{{Inflation by non-minimal coupling}},
  \href{https://doi.org/10.1088/1475-7516/2008/08/009}{\textit{JCAP} {\bfseries
  08} (2008) 009}, [\href{https://arxiv.org/abs/0801.1722}{{\ttfamily
  0801.1722}}].

\bibitem{Planck:2013jfk}
{\scshape Planck} collaboration, P.~A.~R. Ade et~al., \textit{{Planck 2013
  results. XXII. Constraints on inflation}},
  \href{https://doi.org/10.1051/0004-6361/201321569}{\textit{Astron.
  Astrophys.} {\bfseries 571} (2014) A22},
  [\href{https://arxiv.org/abs/1303.5082}{{\ttfamily 1303.5082}}].

\bibitem{Coleman}
S.~Coleman and E.~Weinberg, \textit{Radiative corrections as the origin of
  spontaneous symmetry breaking},
  \href{https://doi.org/10.1103/PhysRevD.7.1888}{\textit{Phys. Rev. D}
  {\bfseries 7} (Mar, 1973) 1888--1910}.

\bibitem{Staub:2013tta}
F.~Staub, \textit{{SARAH 4 : A tool for (not only SUSY) model builders}},
  \href{https://doi.org/10.1016/j.cpc.2014.02.018}{\textit{Comput. Phys.
  Commun.} {\bfseries 185} (2014) 1773--1790},
  [\href{https://arxiv.org/abs/1309.7223}{{\ttfamily 1309.7223}}].

\bibitem{Gorda:2018hvi}
T.~Gorda, A.~Helset, L.~Niemi, T.~V.~I. Tenkanen and D.~J. Weir,
  \textit{{Three-dimensional effective theories for the two Higgs doublet model
  at high temperature}},
  \href{https://doi.org/10.1007/JHEP02(2019)081}{\textit{JHEP} {\bfseries 02}
  (2019) 081}, [\href{https://arxiv.org/abs/1802.05056}{{\ttfamily
  1802.05056}}].

\bibitem{Andersen:1998br}
J.~O. Andersen, \textit{{Dimensional reduction of the two Higgs doublet model
  at high temperature}},
  \href{https://doi.org/10.1007/s100520050655}{\textit{Eur. Phys. J. C}
  {\bfseries 11} (1999) 563--570},
  [\href{https://arxiv.org/abs/hep-ph/9804280}{{\ttfamily hep-ph/9804280}}].

\bibitem{Branco:2011iw}
G.~C. Branco, P.~M. Ferreira, L.~Lavoura, M.~N. Rebelo, M.~Sher and J.~P.
  Silva, \textit{{Theory and phenomenology of two-Higgs-doublet models}},
  \href{https://doi.org/10.1016/j.physrep.2012.02.002}{\textit{Phys. Rept.}
  {\bfseries 516} (2012) 1--102},
  [\href{https://arxiv.org/abs/1106.0034}{{\ttfamily 1106.0034}}].

\bibitem{Bandyopadhyay:2020djh}
P.~Bandyopadhyay, S.~Jangid and M.~Mitra, \textit{{Scrutinizing Vacuum
  Stability in IDM with Type-III Inverse seesaw}},
  \href{https://doi.org/10.1007/JHEP02(2021)075}{\textit{JHEP} {\bfseries 02}
  (2021) 075}, [\href{https://arxiv.org/abs/2008.11956}{{\ttfamily
  2008.11956}}].

\bibitem{Workman:2022ynf}
{\scshape Particle Data Group} collaboration, R.~L. Workman and Others,
  \textit{{Review of Particle Physics}},
  \href{https://doi.org/10.1093/ptep/ptac097}{\textit{PTEP} {\bfseries 2022}
  (2022) 083C01}.

\end{thebibliography}\endgroup
\bibliographystyle{Ref}
\end{document}